\documentclass[aps,prd,reprint,superscriptaddress,amsmath,amssymb]{revtex4-2}
\usepackage[dvipsnames]{xcolor}
\usepackage{dcolumn}
\usepackage{graphicx}
\usepackage{hyperref}
\hypersetup{colorlinks=true,linkcolor=blue}

\begin{document}
\title{WIMP Dark Matter in bulk viscous non-standard cosmologies}

\author{Esteban González}
\email{esteban.gonzalez@ucn.cl}
\affiliation{Departamento de Física, Universidad Católica del Norte, Avenida Angamos 0610, Casilla 1280, Antofagasta, Chile}

\author{Carlos Maldonado}
\email{carlos.maldonado@uss.cl}
\affiliation{Facultad de Medicina y Ciencia, Universidad San Sebasti\'an, Puerto Montt, Chile}

\author{N. Stefanía Mite}
\email{nelly.mite@ucn.cl}
\affiliation{Departamento de Física, Universidad Católica del Norte, Avenida Angamos 0610, Casilla 1280, Antofagasta, Chile}

\author{Rodrigo Salinas}
\email{rodrigo.salinas@alumnos.ucn.cl}
\affiliation{Departamento de Física, Universidad Católica del Norte, Avenida Angamos 0610, Casilla 1280, Antofagasta, Chile}

\begin{abstract}
In this paper, we explore an extension of the classical non-standard cosmological scenario in which the new field, $\phi$, which interacts with the radiation component in the early universe, experiences dissipative processes in the form of a bulk viscosity. Assuming an interaction term given by $\Gamma_{\phi}\rho_{\phi}$, where $\Gamma_{\phi}$ accounts for the decay rate of the field and $\rho_{\phi}$ corresponds to its energy density, and a bulk viscosity according to the expression $\xi=\xi_{0}\rho_{\phi}^{1/2}$ in the framework of Eckart's theory, we apply this novel non-standard cosmology to study the parameters space for WIMPs Dark Matter candidate production. This parameter space shows deviations from the classical non-standard cosmological scenario, obtaining new regions to search for this candidate. In particular, for certain combinations of the free parameters, we found large regions in which the model can establish the DM and reproduce the current observable relic density.
\end{abstract}
\maketitle

\section{\label{sec:Introduction}Introduction}
The $\Lambda$CDM model is the most successful model in describing the evolution of the universe and fit the observational cosmological data coming from type Ia supernovae \cite{SupernovaSearchTeam:1998fmf,SupernovaCosmologyProject:1998vns,Scolnic:2021amr}, observation of the Hubble parameter \cite{Moresco:2012jh,Zhang:2012mp,Moresco:2015cya}, baryon acoustic oscillations \cite{SDSS:2005xqv}, cosmic microwave background \cite{WMAP:2012fli,Planck:2018vyg}, among others. Despite that, the model has some lacks such as the nature of Dark Matter (DM) and Dark Energy (DE), where the first one is an unknown, non-baryonic, component of the universe, which is approximately five times more abundant than ordinary matter \cite{Planck:2018vyg}. Some DM candidates naturally arise from theories like Super Symmetry \cite{Jungman:1995df} or string theory \cite{King:2006cu}. In general, DM candidates are classified into three groups of interest: Weakly Interacting Slim Particles (WISPs) \cite{Arias:2012az, Ringwald:2012hr}, Weakly Interacting Massive Particles (WIMPs) \cite{Steigman:1984ac,Bertone:2004pz,Arcadi:2017kky, Roszkowski:2017nbc, Arcadi:2024ukq} and Feeble Interacting Massive Particles (FIMPs) \cite{Bernal:2017kxu, Chu:2011be, Hall:2009bx}. The first group consists of light particles produced through non-thermal mechanism to avoid becoming relativistic. Examples of these particles include Axions, Axions-like particles, and Hidden Photons (or Dark Photons). WIMPs, in contrast, are produced in thermal equilibrium with the Standard Model (SM) bath. As the universe expands, these particles freeze their number via a mechanism known as Freeze-Out, since their interactions become inefficient in comparison with the expansion rate of the universe. These particles were very popular due to the so-called WIMP Miracle, which is able to reproduce the current DM relic density by considering an interaction cross section around the Electro-Weak scale and a mass for the particle around $100$ GeV. In fact, to be consistent with the observations in $\Lambda$CDM, the total thermal averaged annihilation cross-section for this group must be $\langle\sigma v\rangle_0=\text{few} \times 10^{-26}\,\text{cm}^3/\text{s}=\text{few}\times 10^{-9}$ GeV \cite{Steigman:2012nb}. Nevertheless, almost the full parameter space for WIMPs particles is already covered without any positive signal. In this direction arises the FIMPs, which are produced through non-thermal mechanism and never reach the equilibrium with the thermal bath. Hence, these particles freeze their number via a mechanism known as Freeze-In. To avoid these candidates entering in the equilibrium, their interaction must be even weaker than WIMPs, becoming the FIMPs in an elusive particle since their feeble interactions are difficult to detect with the current instrument. Therefore, it is imperative to find new DM candidates, mechanisms of production or different cosmological scenarios.

In $\Lambda$CDM it is assumed that the DM established (froze) its number during a radiation dominated era, which sets the parameters for its search. However, by introducing an additional field ($\phi$) in the early universe, it is possible to modify the expansion rate, generating non-standard cosmological histories. That may result in the DM relic density being established in eras different from radiation dominated, making imprints on its relic abundance. When the $\phi$ state decays into the SM, produce an entropy injection to the SM bath, translating into new parameter space to search these DM particles \cite{Giudice:2000ex,Salati:2002md,Pallis:2004yy,Gelmini:2006pw,Gelmini:2006pq,Randall:2015xza, Tenkanen:2016jic, Hamdan:2017psw, DEramo:2017ecx, DEramo:2017gpl, Visinelli:2017qga, Drees:2018dsj, Bernal:2018ins, Bernal:2018kcw, Poulin:2019omz, Maldonado:2019qmp, Arias:2019uol, Bernal:2019mhf, Cosme:2020mck, Arcadi:2021doo,Arias:2021rer, Bernal:2022wck, Haque:2023yra, Silva-Malpartida:2023yks,Barman:2024mqo,Ghosh:2023tyz, Silva-Malpartida:2024emu, Arcadi:2024tib}. These non-standard cosmological histories can also be generated considering exotic models such as a bi-metric model, exhibiting the same behavior with an entropy injection to the SM bath \cite{Maldonado:2021aze}, and making imprints in the DM production \cite{Maldonado:2023alg, Maldonado:2023zur}. These scenarios are called Non-Standard Cosmologies (NSCs) and bring us new regions in the DM search or re-open windows with parameters space that are discarded in the $\Lambda$CDM model, but which could be allowed in these scenarios.

If DM is experimentally detected, its particle physics properties, such as mass and interaction with the SM, will be reconstructed, including their couplings. However, the production of the DM candidate and the scenario that establishes its number are significant in determining the right current relic density. In this context, if the interactions of the detected DM are consistent with $\langle\sigma v\rangle_0$, the $\Lambda$CDM scenario is favored. On the other hand, if this is not the case, it is imperative to propose alternative cosmological scenarios that might better explain the DM relic density. One possibility to generate new NSC scenarios is the inclusion of non-perfect fluids in the early universe.

In cosmology, all the matter components of the universe are generally described as perfect fluids, providing a good approximation of the cosmic medium. Nevertheless, perfect fluids come from the thermodynamic equilibrium, where their entropy does not increase, and their dynamics are reversible. When non-perfect fluids are considered, effects like viscosity appear, which provide a more realistic description of these cosmic fluids \cite{Maartens:1996vi}, and are relevant in many cosmological processes like reheating of the universe, decoupling of neutrinos from the cosmic plasma, nucleosynthesis, among others. On the other hand, viscosity can also be present in several astrophysical mechanisms as the collapse of radiating stars into neutron stars or black holes and in the accretion of matter around neutron stars or black holes \cite{Maartens:1996vi}. Following this line, viscous fluids must be described by some relativistic thermodynamical approach to non-perfect fluids like Eckart's \cite{Eckart:1940zz,PhysRev.58.919} or Israel-Stewart's theories \cite{1977RSPSA.357...59S,1979RSPSA.365...43I}. Despite the fact Eckart's theory is a non-causal theory \cite{Israel:1976tn}, it is widely investigated in the literature due to its mathematical simplicity in comparison with the full Israel-Stewart theory and became the starting point to shed some light on the behavior of the dissipative effects since the Israel-Stewart’s theory is reduced to the Eckart’s theory when the relaxation time for transient viscous effects is negligible \cite{Maartens:1995wt}.

It is known from hydrodynamics that there are two types of viscosity, the shear and bulk viscosity. While the shear viscosity must be important in some scenarios \cite{Floerchinger:2014jsa}, we will focus our study on the bulk viscosity, which can arise due to the existence of mixtures in the universe. In this sense, in a single fluid description, the universe as a whole can be characterized by the particle number density $n$ as $n=n_{1}+...+n_{i}$. So, the simple assumption of different cooling rates in the expanding mixture can lead to a non-vanishing viscous pressure \cite{Zimdahl:1996fj}. Another explanation for bulk viscous origin is due to DM decaying into relativistic particles, emerging naturally dissipative effects in the cosmic fluid \cite{Wilson:2006gf,Mathews:2008hk}. Even more, bulk viscosity can appear in the Cold DM (CDM) fluid due to the energy transferred from the CDM fluid to the radiation fluid \cite{Hofmann:2001bi}, and it may also manifest as a component within a hidden sector, reproducing several observational properties of disk galaxies \cite{Foot:2014uba,Foot:2016wvj}. From the point of view of cosmological perturbations, it has been shown that viscous fluid dynamics provides a simple and accurate framework for extending the description of cosmological perturbations into the nonlinear regime \cite{Blas:2015tla}. Finally, the new era of gravitational wave detectors has also opened the possibility of detecting dissipative effects in DM and DE through the dispersion and dissipation experimented by these waves propagating in a non-perfect fluid \cite{Goswami:2016tsu}. As a matter of fact, bulk viscosity could contribute significantly to the emission of gravitational waves in neutron star mergers \cite{Alford:2017rxf}.

The effects of bulk viscosity in the matter components of the universe have been extensively studied in the literature, as the existence of a viscous DE component \cite{Nojiri:2004pf,Capozziello:2005pa,Nojiri:2005sr,Cataldo:2005qh,Brevik:2006wa,Cruz:2016rqi}. However, the most common case is to consider a DM that experiences dissipative processes during its cosmic evolution \cite{Velten:2012uv,Acquaviva:2014vga,Cruz:2017bcv,Cruz:2018yrr,Cruz:2022wme,Cruz:2022zxe,Gomez:2022qcu,Cruz:2023dzn}, which can describe the recent acceleration expansion of the universe without the inclusion of a DE component (unified DM models) \cite{Fabris:2005ts,Avelino:2008ph,Li:2009mf,Avelino:2010pb,Hipolito-Ricaldi:2010wrq,Velten_2011,Gagnon:2011id,Bruni:2012sn,Cruz:2017lbu,Cruz:2018psw,Cruz:2019uya}. A dissipative stiff matter fluid was previously studied in \cite{Mak:2003gw} in the full Israel-Stewart theory. Also, bulk viscous DM has been studied in the context of inflation \cite{Padmanabhan:1987dg,Barrow:1988yc,Maartens:1995wt,Maartens:1996dk}, interacting fluids \cite{Zimdahl:1996fj,Avelino:2013wea,Hernandez-Almada:2020ulm}, and modified gravity \cite{Brevik:2005ue,Brevik:2006wa}. Even more, the viscous effect has been studied in the context of singularities, like Big Rip and Little Rip, for a classical and quantum regime \cite{Nojiri:2004pf,Nojiri:2005sr,Brevik:2005bj,Brevik:2006wa,Brevik:2008xv,Brevik:2010okp,Brevik:2010jv,Brevik:2011mm,Contreras:2015ooa,Contreras:2018two,Cruz:2021knz}. Other scenarios of interest can be found in Refs. \cite{Barta:2019tpv,BravoMedina:2019han}, where the role of bulk viscosity is studied in the radial oscillation of relativistic stars and their cosmological implications for universes filled with Quark-Gluon plasma, respectively. Last but not least, bulk viscosity was also considered as an alternative to alleviate some recent tensions in the $\Lambda$CDM model. For example, a decaying scenario for DM increases the expansion rate relative to $\Lambda$CDM and such behavior provides an alleviation for the $H_{0}$ and $\sigma_{8}$ tensions \cite{Pandey:2019plg}. In Refs. \cite{Yang:2019qza,DiValentino:2021izs,Normann:2021bjy}, bulk viscous effects are explored as a viable alternative to relieve the $H_{0}$ tension. For an extensive review of viscous cosmology in the early and late time universe see Ref. \cite{Brevik:2017msy}.

This paper aims to study how dissipative effects in the form of bulk viscosity left imprints in WIMPs DM production and its relic density in a non-standard cosmology. In particular, we go further than the classical NSCs scenarios, in which the early universe is dominated by two interacting fluids, namely the new field $\phi$ and radiation, by considering that $\phi$ experiences dissipative processes during their cosmic evolution in the form of a bulk viscosity. Working in the framework of Eckart's theory for non-perfect fluids, we compare the NSC scenario with their bulk viscous counterpart, for an specific election of the bulk viscosity. Also, we study the parameter space that can reproduce the current DM relic density varying both, the model and DM parameters, for this novel NSC.

This paper is organized as follows: In Section \ref{sec:originalmodel}, we briefly review the original NSC scenario. Their applicability to WIMP DM is studied in Section \ref{sec:WIMPS}. In section \ref{sec:Viscousmodel}, we describe the bulk viscous extension to the original NSC model, being compared the two models in Section \ref{sec:Comparison}. The parameter space for DM production that leads to the current relic density is analyzed in Section \ref{sec:Parameters}. Finally, in section \ref{sec:Conclusions}, we present some conclusions and future remarks. Through this paper, we consider $c=1$ units.

\section{\label{sec:originalmodel}Non-standard cosmologies}
In the $\Lambda$CDM model, the total energy density budget of the universe at early times is composed of radiation ($\rho_{\gamma}$) and DM ($\rho_{\chi}$), with a negligible cosmological constant in comparison with the other fluids. Following Refs. \cite{Giudice:2000ex, Hamdan:2017psw, DEramo:2017ecx, DEramo:2017gpl, Visinelli:2017qga, Drees:2018dsj, Bernal:2018ins, Bernal:2018kcw, Maldonado:2019qmp, Arias:2019uol, Bernal:2019mhf, Arias:2021rer,Bernal:2022wck, Silva-Malpartida:2024emu}, a straightforward manner to produce NSCs scenarios is adding an extra field $\phi$ in the early universe, with an associated energy density $\rho_{\phi}$,  which will decay in SM plasma. The Friedman equation and the continuity equation in this scenario are
\begin{eqnarray}
    &&3H^2=\frac{\rho_t}{M_p^2},\label{hub}\\
    &&\dot{\rho}_t+3H(\rho_{t}+p_{t})=0,\label{cons}
\end{eqnarray}
where ``dot'' accounts for the derivative with respect to the cosmic time, $H\equiv \dot{a}/a$ is the Hubble parameter, with $a$ the scale factor, and $M_p=2.48\times10^{18}$ GeV is the reduced mass Planck. The total energy density and pressure of the universe are $\rho_{t}=\rho_{\gamma}+\rho_{\phi}+\rho_{\chi}$ and $p_{t}=p_{\gamma}+p_{\phi}+p_{\chi}$, respectively. Also, in $\Lambda$CDM and NSCs scenarios, the DM component is included through the following Boltzmann equation, which accounts for its number density $n_{\chi}$
\begin{equation}
    \dot{n_\chi}+3H n_\chi=-\langle\sigma v\rangle\left(n_\chi^2-n_{eq}^2\right) \label{boltzdm},
\end{equation}
where $\langle\sigma v\rangle$ is the total thermal averaged annihilation cross-section and $n_\text{eq}$ is the equilibrium number density defined as $n_\text{eq}=m_\chi^2\, T\, K_2(m_\chi/T)/\pi^2$, with $K_2$ the Bessel function of second kind, $m_\chi$ the DM mass and $T$ the temperature of the universe. The DM energy density is related to its mass by $\rho_\chi=m_\chi\, n_\chi$.

To consider the relativistic degrees of freedom it is necessary to incorporate the entropy density ($s$) of the universe defined through the radiation energy density as
\begin{equation}
s=\frac{\rho_\gamma+p_\gamma}{T} =\frac{2\pi^2}{45}g_{\star s}(T)T^3 \label{entropy},
\end{equation}
where $g_{\star s}(T)$ are the degrees of freedom that contribute to the entropy density. Therefore, assuming an interaction between the new field $\phi$ and the radiation component, we can obtain from Eq. \eqref{cons} and \eqref{entropy} the following equations
\begin{eqnarray}
    \dot{s}+3Hs&=& \frac{\Gamma_\phi \rho_\phi}{T}, \label{dots}\\ \dot{\rho_\phi}+3\left(\omega+1\right)H\rho_\phi&=&-\Gamma_\phi \rho_\phi, \label{phi}
\end{eqnarray}
where we consider for the field a barotropic equation of state (EoS) of the form $p_{\phi}=\omega\rho_{\phi}$, with $\omega$ the barotropic index. In these equations, $\Gamma_{\phi}\rho_{\phi}$ is the interaction term, where $\Gamma_\phi$ accounts for the decay rate of the new field, and is the most simple (and most studied) case as a NSC scenario. Note that the energy density for the DM candidate can be neglected in Eq. (\ref{hub}) and decoupled from the problem due to the subdominant contribution compared with $\phi$ and radiation. Finally, we can rewrite Eq. \eqref{dots} in terms of the temperature using Eq. \eqref{entropy}, being obtained
\begin{equation} \dot{T}=\left(-HT+\frac{\Gamma_\phi \rho_\phi}{3s}\right)\left(\frac{dg_{\star s}(T)}{dT}\frac{T}{3g_{\star s}(T)}+1\right)^{-1}. \label{temp}
\end{equation}
The latter can be related to the energy density of radiation as $\rho_\gamma=\pi^2g_{\star}(T)\, T^4/90$, with $g_\star (T)$ the degrees of freedom that contribute to the plasma energy density.

Approximated analytical solutions for Eqs. \eqref{dots} and \eqref{phi} can be straightforwardly obtained according to Ref. \cite{Arias:2021rer} as
\begin{eqnarray}
    \frac{\rho_\gamma(a)}{\rho_{\gamma,\text{ini}}}&\simeq&\left(\frac{a_\text{ini}}{a}\right)^4+\frac{2}{8-3(\omega+1)}\frac{\Gamma_\phi}{H_\text{ini}}\left(\frac{a_\text{ini}}{a}\right)^{\frac{3(\omega+1)}{2}},\label{anrad}\\
    \frac{\rho_\phi(a)}{\rho_{\phi,\text{ini}}}&\simeq&\left(\frac{a_\text{ini}}{a}\right)^{3(\omega+1)}-\frac{2}{3(\omega+1)}\frac{\Gamma_\phi}{H_\text{ini}}\left(\frac{a_\text{ini}}{a}\right)^{\frac{3(\omega+1)}{2}},\label{anphi}
\end{eqnarray}
where the subscript ``$\text{ini}$'' correspond to the respective energy density evaluated at the initial temperature $T_\text{ini}=m_\chi$. These solutions are only valid when $\omega\neq-1$, value that indeterminate Eq. \eqref{anphi}, and show the behaviour of both energy densities as a function of the scale factor. Note that the evolution of $\rho_{\phi}$ is mainly dominated by two terms. The first one is the usual expression for the energy density for a barotropic fluid and, the other one, corresponds to a modification due to its interaction with the radiation component. The first one decreases more than the second term as the scale factor grows and, therefore, the fluid vanishes as the universe is expanding. It is important to note that this is an approximated solution and, hence, the possibility of $\rho_{\phi}<0$ is only due to the approximation made. In the special case in which $\omega=-1$, the approximated analytical solution for $\phi$ can be read as
\begin{equation}
    \frac{\rho_\phi(a)}{\rho_{\phi,\text{ini}}}\simeq 1+\frac{\Gamma_\phi}{H_\text{ini}}\ln \left(\frac{a_\text{ini}}{a}\right), \label{anphi-1}
\end{equation}
where, in this case, the energy density of the extra field does not decay and, instead, increases. This same behaviour is obtained for a phantom fluid in which $\omega<-1$, as it can bee seen from Eq. \eqref{anphi}.

A remarkable point about the inclusion of this extra field $\phi$ is the possibility to influence in the inflation and reheating epochs, as it was studied in \cite{Kofman:1997yn, Ackerman:2010he, Garcia:2020eof, Garcia:2020wiy, Bernal:2022wck, Silva-Malpartida:2023yks, Haque:2024zdq}. In the latter, it was considered that the inflationary field can be described by this field $\phi$ incorporated in the NSC scenario and rolling down to its minimum to generate the reheat epoch. On the other hand, there are several cases studied in which the extra field $\phi$ generate a reheating epoch but not coming from an inflationary field \cite{Gelmini:2006pw, Gelmini:2006pq, Visinelli:2017qga, Maldonado:2019qmp}. Neverteless, it is important to highlight that the NSCs scenarios must not modify the Big Bang Nucleosynthesis (BBN) epoch due to the precise observations posterior to this era, which assume the $\Lambda$CDM model \cite{Sarkar:1995dd, Hannestad:2004px, DeBernardis:2008zz}. So, this new field must decay before the beginning of BBN, i.e, when $T_\text{end}\gtrsim T_{BBN}\sim 4$ MeV, with $T_\text{end}$ the temperature of $\phi$ decays (in the case of reheating the same condition must be fulfilled, considering the temperature of reheat $T_\text{Rh}$ instead of $T_\text{end}$).

One way to estimate when a particle goes out of the thermal bath is to analyze if the interaction of the particles is efficient enough to maintain them in equilibrium or if the expansion rate of the universe suppresses their interactions. Therefore, in the limit $H=\Gamma_\phi$, the $\phi$ particle had fully decayed and the standard $\Lambda$CDM cosmology is recovered, relating the temperature of decay with the decay rate as
\begin{equation}
    T_\text{end}^4= \frac{90}{\pi^2 g_\star(T_\text{end})}M_p^2 \Gamma_\phi^2. \label{tend}
\end{equation}

\begin{figure}
\centering
\includegraphics[width=0.48\textwidth]{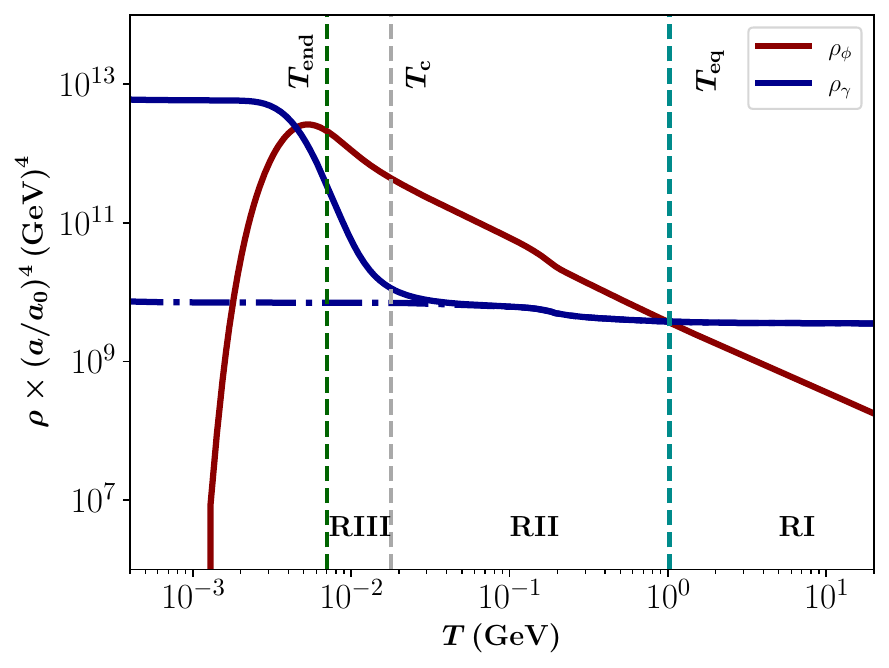}
\caption{Evolution of $\rho\times (a/a_{0})^{4}$ as a function of the temperature $T$ for $\kappa=10^{-2}$, $T_\text{end}=7\times 10^{-3}$ GeV, $m_\chi=100$ GeV, and $\omega=0$. The solid line corresponds to the NSC scenario and the dashed-dotted line to the standard $\Lambda$CDM scenario, while the red line represents the new field $\phi$ and the blue line the radiation component. Also, we depict the values of $T_\text{eq}$ (cyan), $T_\text{c}$ (grey), and $T_\text{end}$ (green).}
\label{figenergydensitystd}
\end{figure}

The inclusion of this new field has some remarkable points, namely $T_\text{eq}$,  $T_\text{c}$, and $T_\text{end}$. The first one corresponds to the temperature when $\rho_\phi=\rho_\gamma$, i.e., the point where $\phi$ starts to dominate over radiation. $T_\text{c}$  corresponds to the temperature at which the decays of this new field begin to become significant in the entropy injection for $\rho_\gamma$. Finally, $T_\text{end}$ is the temperature when $\phi$ decays, as we mentioned before. With these identifications, we can define three regions of interest: RI) $T_\text{eq}<T$, RII) $T_\text{eq}>T>T_\text{c}$, and RIII) $T_\text{c}>T>T_\text{end}$. A fourth region (RIV) that is not of interest to us can be considered where $T<T_\text{end}$, which corresponds to the standard $\Lambda$CDM scenario where the new field has fully decayed. Therefore, the NSC scenario will be characterized by the parameters $\kappa\equiv\rho_{\phi,\text{ini}}/\rho_{\gamma,\text{ini}}$, the barotropic index $\omega$, and the temperature $T_\text{end}$. As an example, in Fig. \ref{figenergydensitystd} we depict a NSC scenario for $\kappa=10^{-2}$, $T_\text{end}=7\times 10^{-3}$ GeV, $\omega=0$, and $m_\chi=100$ GeV.

\begin{figure}
\centering
\includegraphics[width=0.48\textwidth]{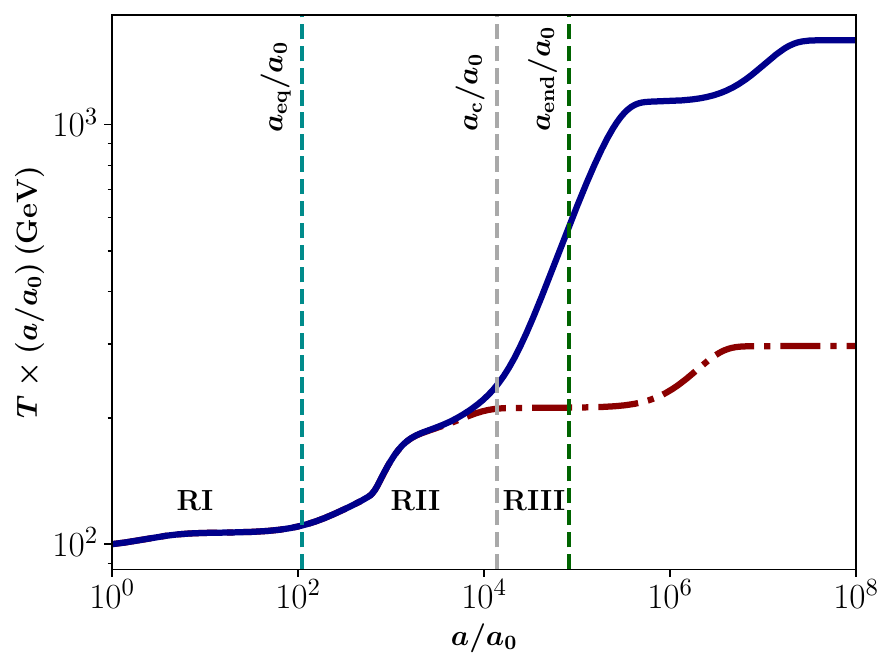}
\caption{Evolution of $T\times (a/a_{0})$ as a function of $a/a_0$ for $\kappa=10^{-2}$, $T_\text{end}=7\times 10^{-3}$ GeV, $m_\chi=100$ GeV, and $\omega=0$. The solid blue line corresponds to the NSC scenario and the dashed-dotted red line to the standard $\Lambda$CDM scenario. Also, we depict the values of $T_\text{eq}$ (cyan), $T_\text{c}$ (grey), and $T_\text{end}$ (green).}
\label{figtempstd}
\end{figure}

Considering constant degrees of freedom, it can be shown that the energy density of the field $\phi$ goes as $\rho_\phi\propto a^{-3(\omega+1)}$. On the other hand, for regions I and II, the temperature goes as $T\propto a^{-1}$ when $T>T_\text{c}$; while in region III, the temperature goes as $T\propto a^{-3(\omega+1)/8}$. The temperature takes the usual form $T\propto a^{-1}$ after the full decay of $\phi$, as it was shown in Eq. \eqref{anphi}, recovering the standard $\Lambda$CDM cosmology. This behavior of the temperature can be seen in Fig. \ref{figtempstd}, for a NSC scenario with $\kappa=10^{-2}$, $T_\text{end}=7\times 10^{-3}$ GeV, $\omega=0$, and $m_\chi=100$ GeV. It is important to note that the fluctuations in the temperature and radiation energy density are produced by the full numerical integration, including the degrees of freedom for entropy and radiation.

\subsection{\label{sec:WIMPS}WIMPs in non-standard cosmologies}
The WIMPs are thermally produced in the early universe, being in equilibrium with the thermal bath. They are very popular DM candidates due to the so-called WIMP miracle where, to reproduce the observations in $\Lambda$CDM, the total thermal averaged annihilation cross-section must be $\langle\sigma v\rangle_0=\text{few} \times 10^{-9}$ GeV$^{-2}$ to obtain the current DM relic density. Hence, larger values of $\langle\sigma v\rangle$ remain the particles in the thermal bath and, therefore, when its number is frozen it produces an under-abundance of DM relic density which can be alleviated with a multi-component DM. On the other hand, if $\langle\sigma v\rangle<\langle\sigma v\rangle_0$, the DM particles go out of equilibrium quickly and the over-abundance of DM relic density forbids those values for their interaction.

To obtain the DM relic density it is useful to define the Yield of DM as $Y\equiv n_\chi/s$ and the dimensionless quantity $x\equiv m_\chi/T$. An analytical solution for Eq. \eqref{boltzdm} can be obtained in the limit $Y\gg Y_\text{eq}$, giving
\begin{equation}
    Y\propto \frac{1}{m_\chi J(x_\text{fo})}, \label{yield}
\end{equation}
with $J=\int_{x_\text{fo}}^\infty x^{-2}\langle\sigma v\rangle(x) dx$ an integral depending on $x$, where $x_\text{fo}$ correspond to the time at which the DM particle goes out of the equilibrium and freeze its number. Note that if the total thermal averaged annihilation cross-section is constant, then the integral turns out to $\langle\sigma v\rangle/x_\text{fo}$. The latter expression shows that if $\langle\sigma v\rangle$ grows, the DM Yield decreases (and vice-versa). An expression for the quantity $x_\text{fo}$ is obtained when the DM can not compete with the universe expansion, i.e., when $H=\Gamma=n_\text{eq}\langle\sigma v\rangle$, from which appears a transcendental equation for $x_\text{fo}$.

\begin{figure}
\centering
\includegraphics[width=0.48\textwidth]{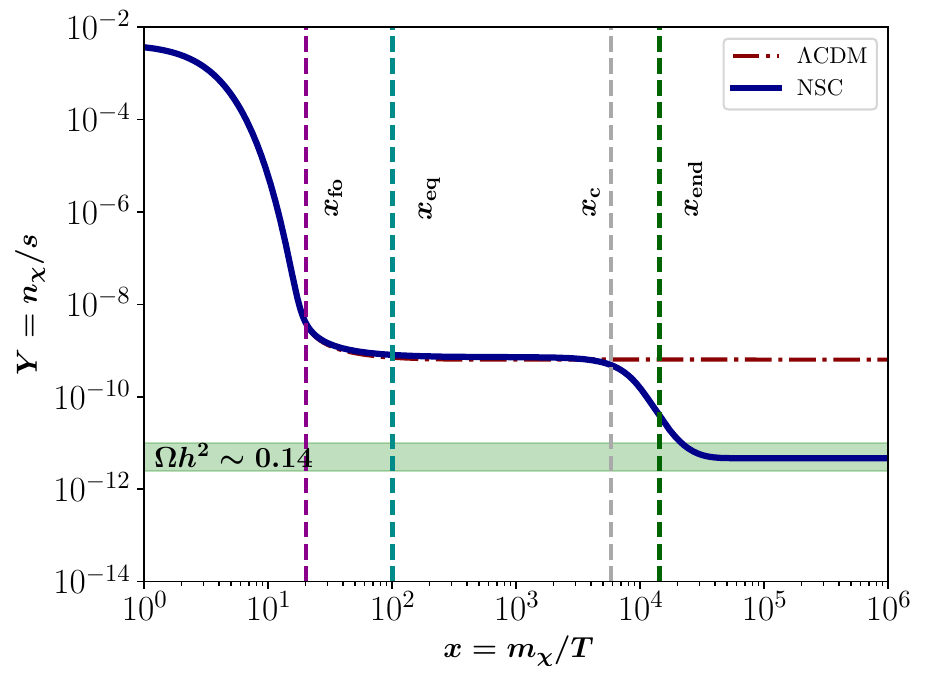}
\caption{Comparison in the yield production in the NSC (solid blue line) and the $\Lambda$CDM scenario (dashed-dotted red line) for $m_\chi=100$ GeV, $\langle\sigma v\rangle=10^{-11}$ GeV$^{-2}$, $\omega=0$, $\kappa=10^{-2}$, and $T_\text{end}=7\times 10^{-3}$ GeV. The dashed lines correspond to $x_\text{eq}$ (cyan), $x_\text{c}$ (grey), $x_\text{end}$ (green), and $x_\text{fo}$ (magenta). The green zone represents the current DM relic density, observing that the parameter space considered in this case gives us the right amount of DM in the NSC.}
\label{figyieldR1}
\end{figure}

\begin{figure}
\centering
\includegraphics[width=0.48\textwidth]{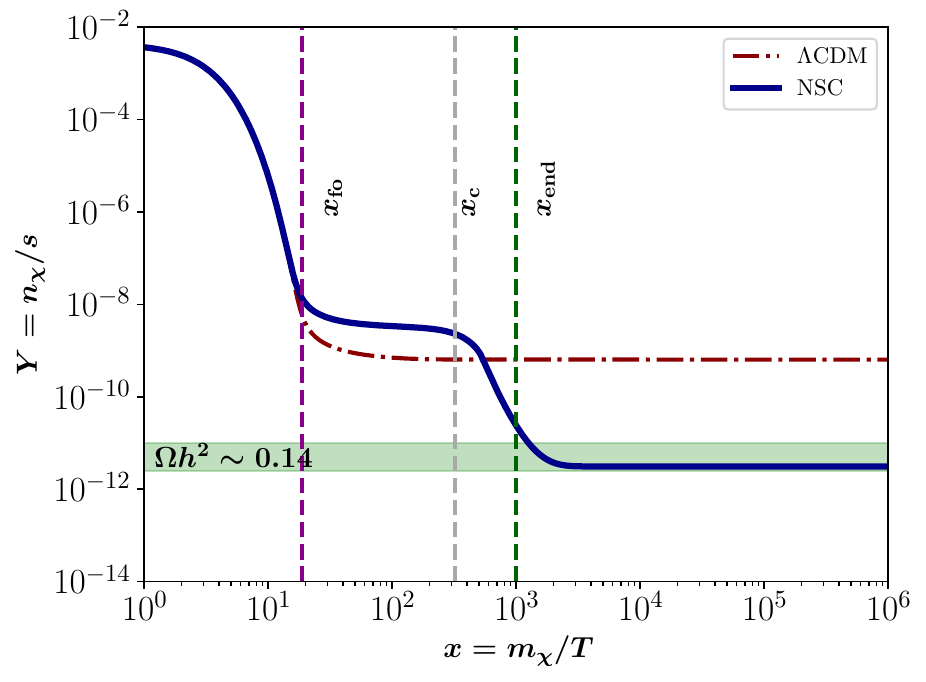}
\caption{Comparison in the yield production in the NSC (solid blue line) and $\Lambda$CDM scenario (dashed-dotted red line) for $m_\chi=100$ GeV, $\langle\sigma v\rangle=10^{-11}$ GeV$^{-2}$, $\omega=0$, $\kappa=1$, and $T_\text{end}=0.1$ GeV. The dashed lines correspond to  $x_\text{c}$ (grey), $x_\text{end}$ (green), and $x_\text{fo}$ (magenta). The green zone represents the current DM relic density, observing that the parameter space considered in this case gives us the right amount of DM in the NSC.}
\label{figyieldR2}
\end{figure}

\begin{figure}
\centering
\includegraphics[width=0.48\textwidth]{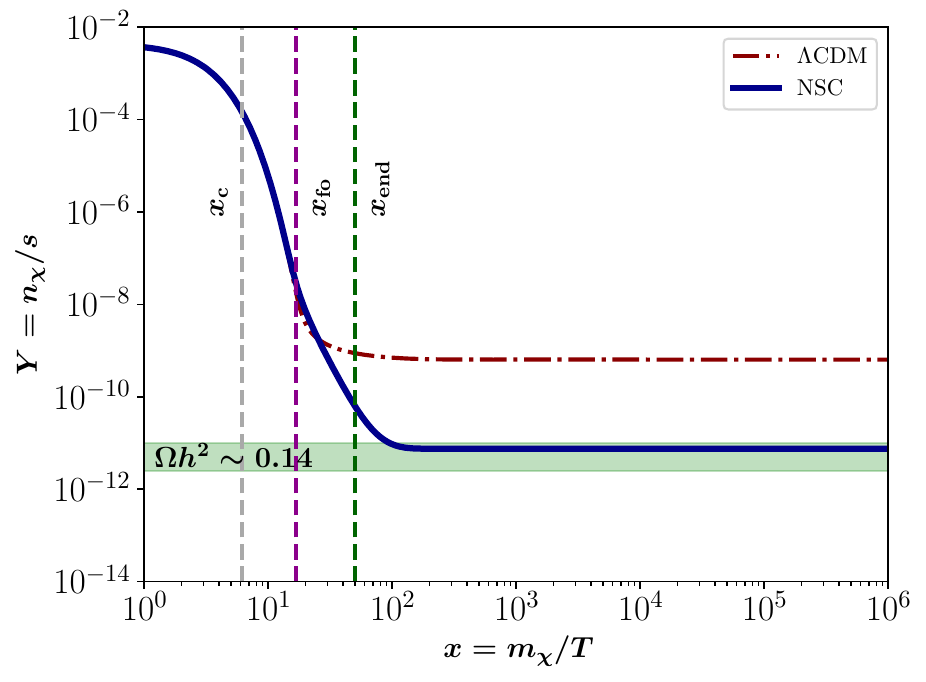}
\caption{Comparison in the yield production in the NSC (solid blue line) and $\Lambda$CDM scenario (dashed-dotted red line) for $m_\chi=100$ GeV, $\langle\sigma v\rangle=10^{-11}$ GeV$^{-2}$, $\omega=0$, $\kappa=10^3$, and $T_\text{end}=2$ GeV. The dashed lines correspond to $x_\text{c}$ (grey), $x_\text{end}$ (green), and $x_\text{fo}$ (magenta). The green zone represents the current DM relic density, observing that the parameter space considered in this case gives us the right amount of DM in the NSC.}
\label{figyieldR3}
\end{figure}

In the NSC scenario, the new field produces a boost in the radiation energy density (and in temperature) that can be parameterized by an entropy injection at the time where $\phi$ starts its decay. This is defined as the entropy density before and after the field $\phi$ decays, i.e., $D\equiv s(T_\text{end})/s(m_\chi)=\left(T_\text{end}/m_\chi\right)^3$. This entropy injection dilutes the DM relic density, considering that the DM Yield depends on entropy density and, therefore, an increment in the entropy density of radiation generates lower values of DM yield. This means that the parameters $m_\chi$ and $\langle\sigma v\rangle$ that overproduce DM can be allowed in this NSC scenario. Hence, the DM relic can be established in four cases, which are related to the four regions mentioned before:
\begin{itemize}
    \item RI: This region exists only for $\kappa<1$ ($\rho_{\gamma,\text{ini}}>\rho_{\phi,\text{ini}}$) with a Hubble parameter of the form $H\sim\sqrt{\rho_\gamma/3M_p^2}\propto T^2$, i.e., approximately the same Hubble parameter as the standard $\Lambda$CDM cosmology. This case is shown in Fig. \ref{figyieldR1}, for a NSC with $\kappa=10^{-2}$, $T_\text{end}=7\times 10^{-3}$ GeV, $\langle\sigma v\rangle=10^{-11}$ GeV$^{-2}$, $m_\chi=100$ GeV, and $\omega=0$. In this case, the DM freezes its number at the same time in the NSC and in the standard $\Lambda$CDM case, overproducing the observed relic density. Nevertheless, when $\phi$ starts to decay, the DM is diluted by the entropy injection, reproducing the current DM relic density. This allows the parameters $\left(m_\chi,\,\langle\sigma v\rangle\right)=\left(100\, \text{GeV},\,10^{-11}\, \text{GeV}^{-2}\right)$, which were discarded in the $\Lambda$CDM scenario.
    \item RII: In this case, $\rho_\phi$ starts to dominate over $\rho_\gamma$, but the decay of the field is not efficient enough to change the radiation energy density. However, the expansion rate of the universe is dominated by $\phi$ and the Hubble Parameter can be approximated as $H\sim\sqrt{\rho_\phi/3M_p^2}\propto T^{3(\omega+1)/2}$. In Fig. \ref{figyieldR2} it is shown the Yield of DM for $\kappa=1$, $T_\text{end}=0.1$ GeV, $\omega=0$, $m_\chi=100$ GeV, and $\langle\sigma v\rangle=10^{-11}$ GeV$^{-2}$. Note that the Freeze out for the $\Lambda$CDM cosmology happens after the NSC case due to the different rates of expansion of the universe. Nevertheless, after the entropy injection ($\phi$ decays), the DM Yield reproduces the current relic observable in the NSC scenario.
    \item RIII: In this region, $\phi$ is still the dominant fluid of the universe, injecting entropy to the SM bath due to its decay. The expansion rate of the universe can be approximated as $H\sim\sqrt{\rho_\phi/3M_p^2}\propto T^4$ for the decaying period, and the entropy injection dilutes the DM relic density as it shown in Fig. \ref{figyieldR3} for $\kappa=10^3$, $T_\text{end}=2$ GeV, $\omega=0$, $m_\chi=100$ GeV, and $\langle\sigma v\rangle=10^{-11}$ GeV$^{-2}$. Again, the DM freezes its number before the NSC scenario in the decay region, compared with the $\Lambda$CDM scenario. Therefore, the entropy injection to the SM ensures that the DM parameters can reproduce the current relic density in the NSC scenario.
    \item RIV: Finally, in this case the field $\phi$ has fully decay and the $\Lambda$CDM model is recovered, i.e, the DM relic is produced out of the NSC and, therefore, we don't obtain modifications in the DM production.
\end{itemize}

\section{\label{sec:Viscousmodel}Bulk viscous non-standard cosmologies}
The equations that govern the evolution of the universe are obtained through the Einstein field equations
\begin{equation}\label{Einstein}
    R_{\mu\nu}-\frac{1}{2}Rg_{\mu\nu}+\Lambda g_{\mu\nu}=\frac{1}{M_{p}^{2}}T_{\mu\nu},
\end{equation}
where $R_{\mu\nu}$ is the Ricci tensor, $R$ the Ricci scalar, $g_{\mu\nu}$ is the metric tensor of the four-dimensional spacetime, and $T_{\mu\nu}$ is the total energy-momentum tensor. In the NSC scenario, as in the classical description of the universe, the total energy budget is described by a perfect fluid, whose respective energy-momentum tensor can be expressed as
\begin{equation}\label{energy-momentum}
T_{\mu\nu}=p_{t}\,g_{\mu\nu}+\left(\rho_{t}+p_{t}\right)u_{\mu}{u_\nu},
\end{equation}
where $u_{\mu}$ correspond to the four-velocity of the fluid element. So, for the spatially flat Friedmann-Lemaitre-Robertson-Walker (FLRW) metric, given by 
\begin{equation}\label{FLRW}
 dl^2=-dt^{2}+a^{2}(t)\left(dr^{2}+r^{2}d\vartheta^{2}+r^{2}\sin^{2}\left(\vartheta\right) d\varphi^{2}\right),
\end{equation}
we obtain the Friedmann Eq. \eqref{hub} and the acceleration equation
\begin{equation}\label{acceleration}
    2\dot{H}+3H^{2}=-p_{t},
\end{equation}
where we have discarded beforehand the cosmological constant because it is negligible in comparison to the other fluids in the epoch of our interest. The continuity Eq. \eqref{cons} is obtained through the expression $\nabla^{\nu}T_{\mu\nu}=0$.

To consider non-perfect fluids in the model, we use, in particular, the framework of relativistic thermodynamic theory out of equilibrium of Eckart, which introduces a small correction $\Delta T_{\mu\nu}$ to Eq. \eqref{energy-momentum} according to the expression $\Delta T_{\mu\nu}=-3H\xi\left(g_{\mu\nu}+u_{\mu}u_{\nu}\right)$ \cite{PhysRev.58.919}, where $\xi$ is the bulk viscosity. In the latter, we have considered that the dissipative fluid doesn't experience heat flow and shear viscosity. Therefore, the energy-momentum tensor takes the form
\begin{equation}\label{Eckartenergy-momentum}
T_{\mu\nu}=P_{\text{eff}}\,g_{\mu\nu}+\left(\rho+P_{\text{eff}}\right)u_{\mu}u_{\nu},    
\end{equation}
where $P_{\text{eff}}=p_{t}+\Pi$, with $\Pi=-3H\xi$ the bulk viscous pressure, and the Eqs. \eqref{cons} and \eqref{acceleration} becomes
\begin{eqnarray}
    &&2\dot{H}+3H^{2}=-p_{t}-\Pi, \label{Dissacceleration} \\
    &&\dot{\rho}_t+3H(\rho_{t}+p_{t}+\Pi)=0, \label{Disscons}
\end{eqnarray}
while Eq. \eqref{hub} remains unchanged. Note that the bulk viscosity affects the evolution of the universe through the bulk viscous pressure. In particular, for an expanding universe, the expression $\Pi=-3H\xi$ is always negative ($\xi>0$ in order to be consistent with the second law of thermodynamics \cite{SWeinberg}) and, therefore, the viscosity leads to an acceleration in the universe expansion, according to Eq. \eqref{Dissacceleration}. In accordance with this approach, an interesting case can be seen in Ref. \cite{Chowdhury:2023jft}, where the interaction term of the classical NSC scenario can be interpreted as a time-dependent dissipation in an effective way.

We aim to study the effects that the bulk viscosity produces in the classical NSC scenario described in Section \ref{sec:originalmodel}. For this purpose, we need to take into account that the division of the total energy budget of the universe into different components is merely a convention since the energy-momentum tensor describes all the fluids components as a whole. Hence, the effective pressure for this NSC is $P_\text{eff}=p_{\gamma}+p_{\phi}+p_{\chi}+\Pi$, where we can make the identification $P_{\text{eff},\phi}=p_{\phi}+\Pi$, i.e., this new field $\phi$ is the fluid that experience dissipative processes during their cosmic evolution, and Eq. \eqref{phi} becomes
\begin{equation}\label{Dissphi}
\dot{\rho_\phi}+3\left(\omega+1\right)H\rho_\phi=-\Gamma_\phi\rho_\phi-3H\Pi ,
\end{equation}
while the other equations of interest in the NSC scenario remain unchanged. In this sense, bulk viscosity can depend, particularly, on the temperature and pressure of the dissipative fluid \cite{SWeinberg}. Therefore, a natural choice for the bulk viscosity of the dissipative fluid is to consider a dependency proportional to the power of their energy density $\xi=\xi_{0}\rho_{\phi}^{1/2}$, where $\xi_{0}=\hat{\xi_{0}}M_{p}$ in order to obtained $\hat{\xi_0}$ as a dimensionless parameter, election that has been widely investigated in the literature. Therefore, the latter expression takes the form
\begin{equation}\label{eqphibulk}
    \dot{\rho}_{\phi}+3(\omega+1)H\rho_{\phi}=-\Gamma_\phi\rho_\phi+9M_{p}\hat{\xi_0}H^{2}\rho_\phi^{1/2}.
\end{equation}
Note that the parameterization chosen for the bulk viscosity has the advantage that, when the field $\phi$ fully decays in SM plasma, the dissipation becomes negligible and we recover the standard $\Lambda$CDM scenario without viscosity.

For the comparison between the classical NSC scenario and their bulk viscous counterpart, we numerically integrate Eqs. \eqref{hub}, \eqref{boltzdm}, \eqref{phi}, \eqref{temp}, and \eqref{eqphibulk}, showing the results in the next subsection.

\subsection{\label{sec:Comparison} Comparison between scenarios}
In this section, we will compare all the features discussed above between the NSC described in Section \ref{sec:originalmodel} and the bulk viscous NSC described in Section \ref{sec:Viscousmodel}. 

\begin{figure}
\centering
\includegraphics[width=0.48\textwidth]{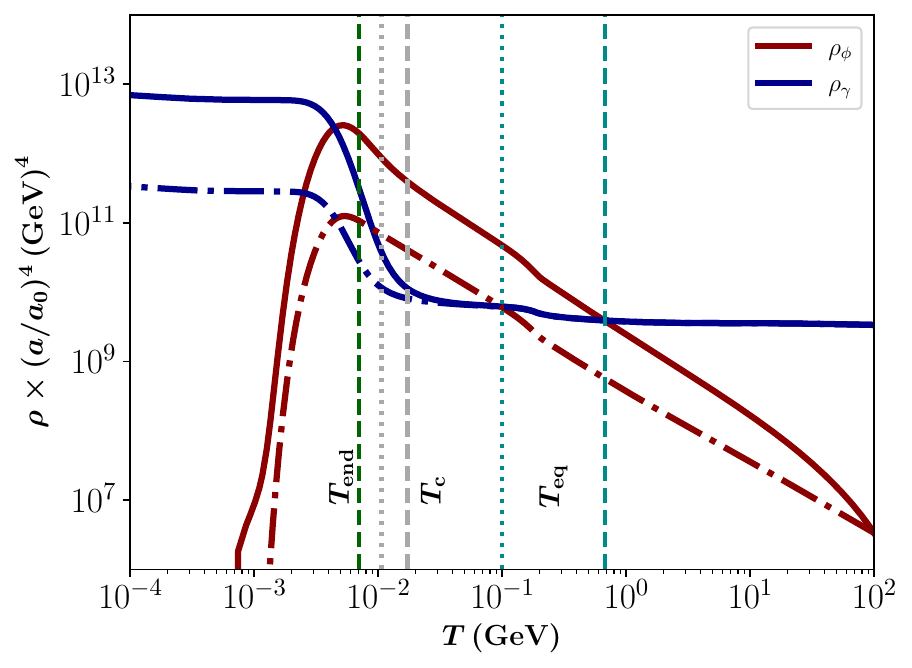}
\caption{Evolution of $\rho\times (a/a_{0})^{4}$ as a function of the temperature $T$ for $\kappa=10^{-2}$, $T_\text{end}=7\times 10^{-3}$ GeV, $m_\chi=100$ GeV, $\hat{\xi_0}=10^{-3}$, and $\omega=0$. The solid and dashed-dotted lines correspond to the NSC with bulk viscosity and the NSC, respectively; while the red and blue lines correspond to the new field $\phi$ and the radiation component, respectively. Also, we depict the values of  $T_\text{eq}$ (cyan), $T_\text{c}$ (grey), and $T_\text{end}$ (green) for the NSC with bulk viscosity (dashed) and the NSC (dotted).}
\label{figenergydensity}
\end{figure}

In Fig. \ref{figenergydensity}, we depict the evolution of $\rho\times (a/a_{0})^{4}$ as a function of the temperature $T$ for both scenarios, considering the values $\kappa=10^{-3}$, $T_\text{end}=7\times 10^{-3}$ GeV, $m_\chi=100$ GeV, $\hat{\xi_0}=10^{-2}$, and $\omega=0$. The solid and dashed-dotted lines correspond to the NSC with bulk viscosity and the NSC, respectively, while the red and blue lines correspond to the new field $\phi$ and the radiation component, respectively. We also present the values of  $T_\text{eq}$ (cyan), $T_\text{c}$ (grey), and $T_\text{end}$ (green) for the NSC with bulk viscosity (dashed) and the NSC (dotted). From the figure, it can be seen a boost in the production of the new field $\phi$ for the bulk viscous NSC in comparison with the NSC case, which leads to a higher increment in the energy density of radiation due to the decay of this viscous state and, therefore, there is a higher entropy injection to the SM bath. Nevertheless, we can see that the behaviour of the fluids as a function of the temperature does not exhibit greater changes between both scenarios and, therefore, we can still ensure that the field $\phi$ becomes negligible for $\omega>-1$ at late times, recovering the usual $\Lambda$CDM model.

\begin{figure}
\centering
\includegraphics[width=0.4\textwidth]{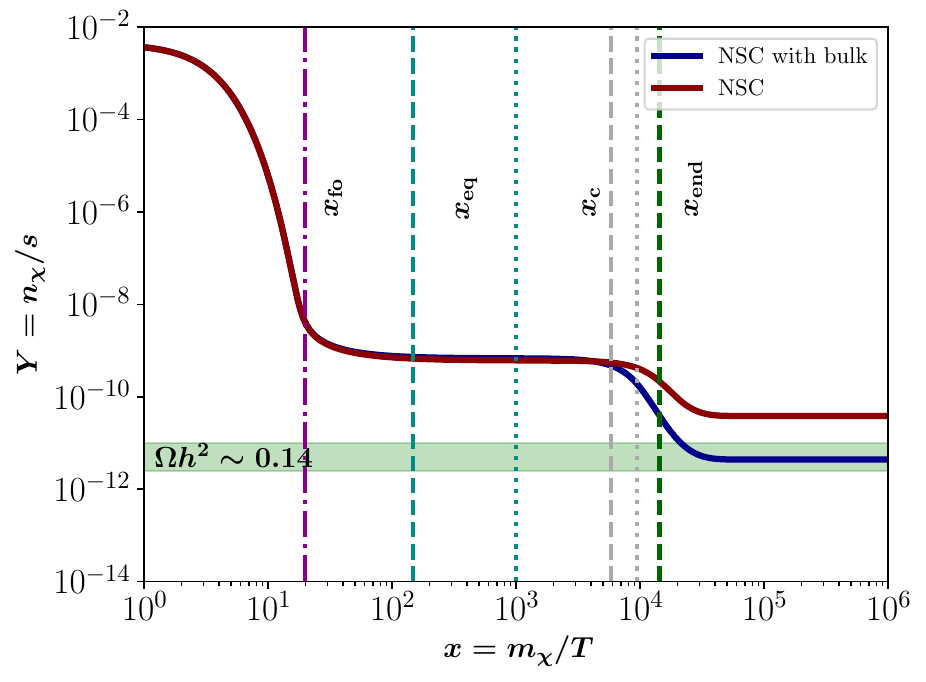}
\caption{Comparison in the yield production in the NSC (red line) and the NSC with bulk viscosity (blue line) for $m_\chi=100$ GeV, $\langle\sigma v\rangle=10^{-11}$ GeV$^{-2}$, $\omega=0$, $\kappa=10^{-3}$ and $T_\text{end}=7\times 10^{-3}$ GeV. The dashed and dotted lines correspond to $x_\text{eq}$ (cyan) and $x_\text{c}$ (grey) for the NSC with bulk viscosity and NSC, respectively. The dashed green line corresponds to $x_\text{end}$, which is the same for both models, and the dashed-dotted line (magenta) is the time when the DM candidate Freeze-Out their number at $x_\text{fo}$. The green zone represents the current DM relic density, observing that the values considered for the parameter space give us the right amount in the NSC with bulk viscosity.}
\label{figyield}
\end{figure}

The Yield DM production is depicted in Fig. \ref{figyield}, showing a comparison between the NSC (red line) and the NSC with bulk viscosity (blue line) for $m_\chi=100$ GeV, $\langle\sigma v\rangle=10^{-11}$ GeV$^{-2}$, $T_\text{end}=7\times 10^{-3}$ GeV, and $\kappa=10^{-3}$. The dashed and dotted-line correspond to $x_\text{eq}$ (cyan) and $x_\text{c}$ (grey) for NSC with and without bulk viscosity, respectively. The green dashed line corresponds to $x_\text{end}$, which is the same for both models, while the magenta dashed-dotted line corresponds to $x_\text{fo}$. The current DM relic density is illustrate in the green strap. In this case the Freeze-Out occurs before $T_\text{eq}$, corresponding to RI, and from $x_\text{fo}$ to $x_\text{c}$, the behavior for both cases is the same in comparison with the $\Lambda$CDM model. Then, the entropy injection begins due to the decay of $\phi$ in both scenarios, which sets the final relic density in agreement with the current observation in the bulk viscous NSC scenario rather than the NSC scenario.

\begin{figure}
\centering
\includegraphics[width=0.4\textwidth]{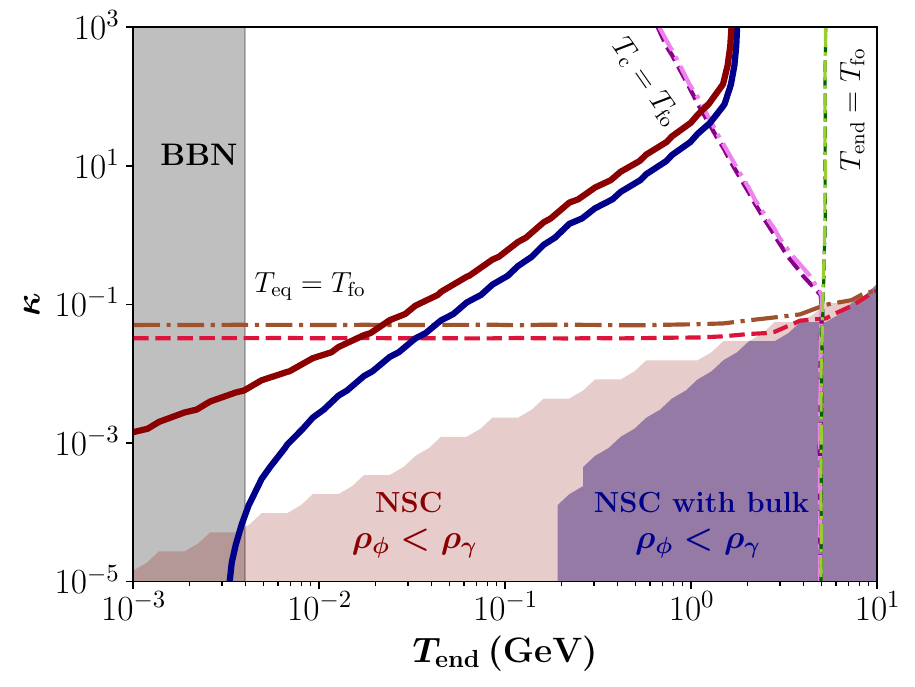}
\caption{Parameter space for the model in $\kappa$ and $T_{end}$ for $\omega=0$, $\hat{\xi_0}=10^{-2}$, $m_\chi=100$ GeV, and $\langle\sigma v\rangle=10^{-11}$ GeV$^{-2}$. The red and blue lines correspond to the parameters that reproduce the current DM relic density for NSC and NSC with bulk viscosity, respectively. The dashed and dashed-dotted lines corresponds to $T_\text{eq}=T_\text{fo}$, $T_\text{c}=T_\text{fo}$, and $T_\text{end}=T_\text{fo}$ for the NSC with and without bulk viscosity, respectively. The grey zone corresponds to the BBN epoch, which starts at $T_\text{BBN}\sim 4\times 10^{-3}$ GeV. The red and blue zones (NSC and NSC with bulk viscosity, respectively) are the parameter space in which the energy density of $\phi$ is lower than radiation. In this case, the model with viscosity allows lower values of $\kappa$ to reproduce the DM relic density. In fact, for almost the same $T_\text{end}$, there exists a considerable range of values for $\kappa$ that reproduces the relic density due to the effect of the viscosity included in $\phi$. }
\label{figkappaTendw0}
\end{figure}

\begin{figure}
\centering
\includegraphics[width=0.4\textwidth]{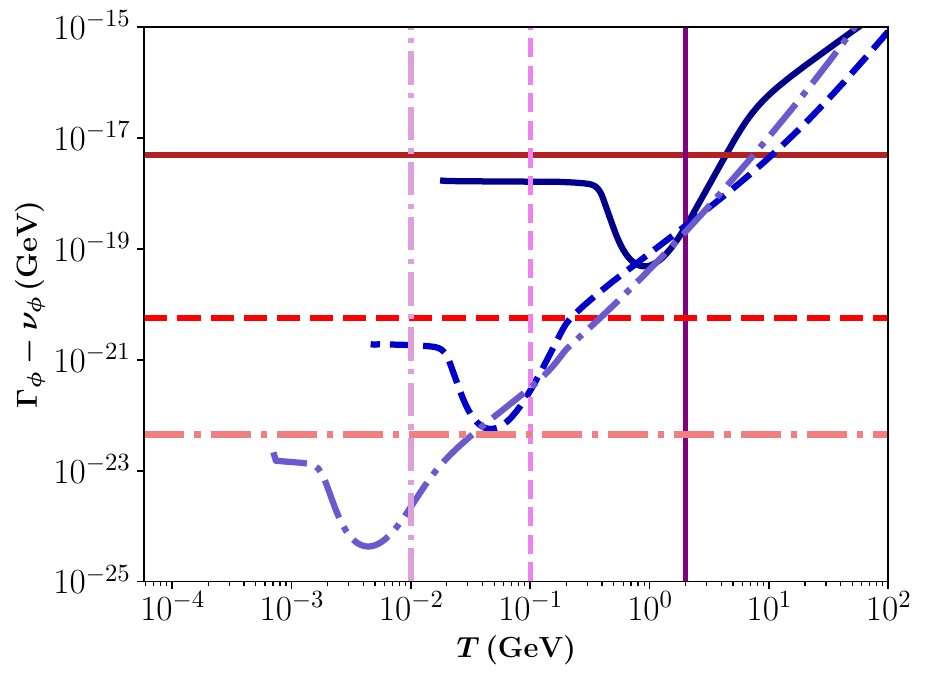}
\caption{Evolution of $\nu_\phi$ and $\Gamma_\phi$ terms for three benchmarks points: RIII with values $\left(T_\text{end},\,\kappa\right)=(2\,\text{GeV},\,100)$ (solid lines), RII with values $\left(T_\text{end},\,\kappa\right)=(10^{-1}\,\text{GeV},\,10^{-1})$ (dashed lines), and RI with values $\left(T_\text{end},\,\kappa\right)=(10^{-2}\,\text{GeV},\,10^{-3})$ (dashed-dotted lines). The figure is depicted for $\omega=0$ and $\hat{\xi_0}=10^{-2}$. The blue, red, and purple color palette correspond to $\nu_\phi$, $\Gamma_\phi$, and $T_\text{end}$, respectively. From this figure we can see that for higher values of $\kappa$ (solid lines) the viscous term is mostly dominated by the decay term, while when the value of $\kappa$ diminishing, the domination of the viscous term is relevant (dashed and dashed-dotted lines).}
\label{figgammanu}
\end{figure}

\begin{figure}
\centering
\includegraphics[width=0.4\textwidth]{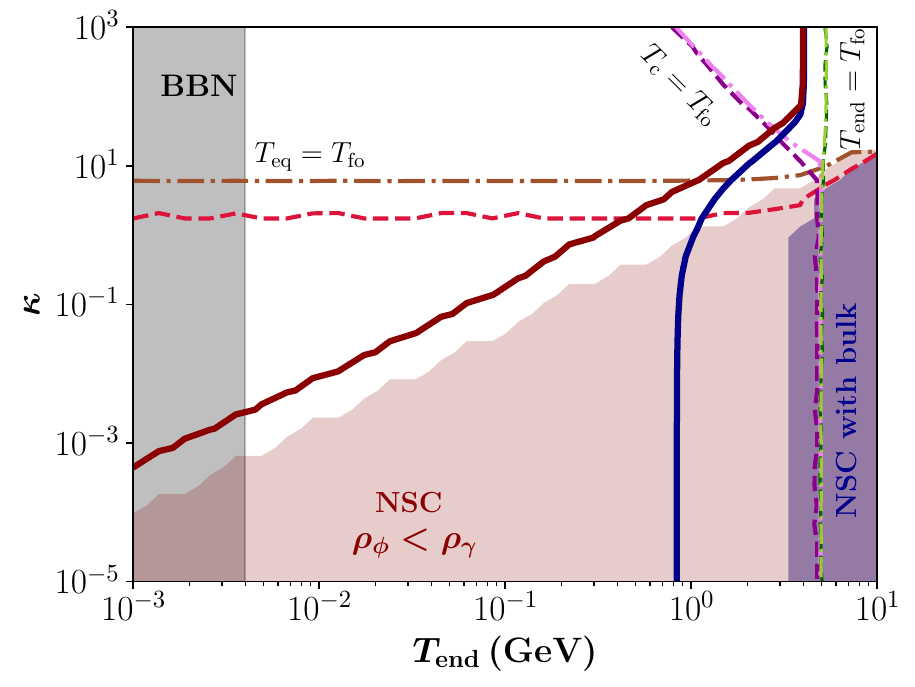}
\caption{Parameter space ($T_\text{end}$, $\kappa$) for the model considering $\omega=-2/5$ and $\hat{\xi_0}=10^{-2}$, and the DM parameters $m_\chi=100$ GeV and $\langle\sigma v\rangle=10^{-11}$ GeV$^{-2}$. The blue and red lines correspond to the parameters space that reproduce the observed DM relic density for a NSC with and without bulk viscosity, respectively. The dashed and dashed-dotted lines correspond to $T_\text{eq}=T_\text{fo}$, $T_\text{c}=T_\text{fo}$, and $T_\text{end}=T_\text{fo}$ for the NSC with and without bulk viscosity, respectively.}
\label{figkappaTendw-25}
\end{figure}

\begin{figure}
\centering
\includegraphics[width=0.4\textwidth]{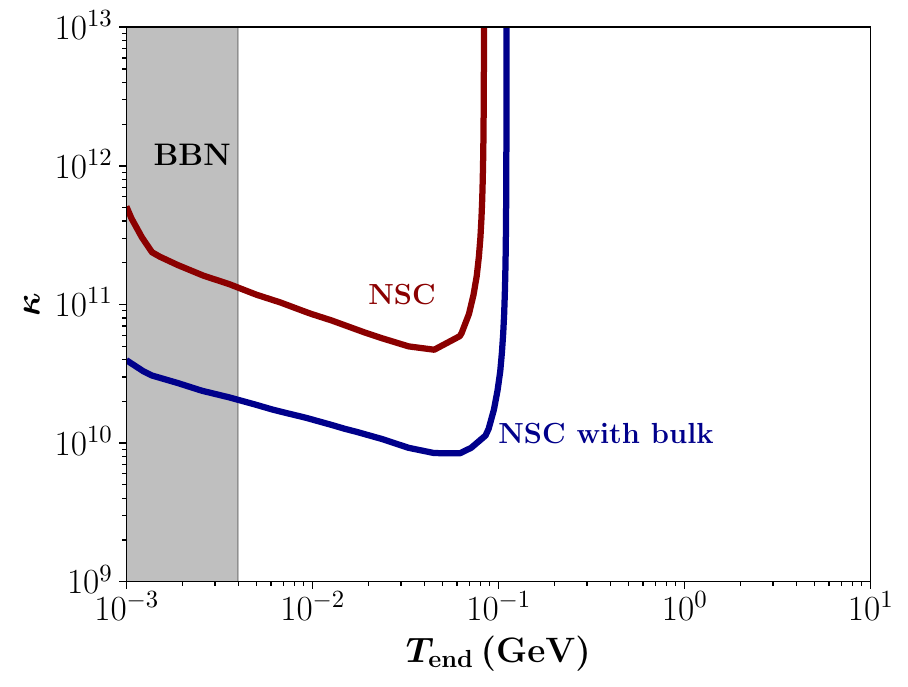}
\caption{Parameter space ($T_\text{end}$, $\kappa$) for the model considering $\omega=2/5$ and $\hat{\xi_0}=10^{-2}$, and the DM parameters $m_\chi=100$ GeV and $\langle\sigma v\rangle=10^{-11}$ GeV$^{-2}$. The blue and red lines correspond to the parameters space that reproduce the observed DM relic density for a NSC with and without bulk viscosity, respectively. Note that the inclusion of viscosity in $\phi$ extend the parameters space of the model to lower values of $\kappa$ (near one orders of magnitude) and slightly larger values of $T_\text{end}$.}
\label{figkappaTendw25}
\end{figure}

A comparison in the models' parameter space $(\kappa,\, T_\text{end})$ that can reproduce the observed DM relic density is shown in Fig. \ref{figkappaTendw0} for the NSC with bulk viscosity (blue line) and the NSC (red line). It is considered the particular case where $\omega=0$, $m_\chi=100$ GeV, and $\langle\sigma v\rangle=10^{-11}$ $\left(\text{GeV}\right)$$^{-2}$. When the DM freezes its number in RIII (higher values of $\kappa$), the two studied cases are similar and reproduce the DM relic density almost for the same parameters. On the other hand, if it is established in RII, the behavior is similar, but there is a deviation from the NSC scenario. Meanwhile the value of $\kappa$ is going lower (RI), the difference between the NSC with and without bulk viscosity is significant. The latter highlight that, for a given value of $T_\text{end}$, large values of $\kappa$ can reproduce the current DM relic density, similarly to the independence of $\kappa$ in RIII. This feature can be explained from Eq. (\ref{eqphibulk}), where we can see that the bulk viscosity has a positive contribution ($\phi$ particles production) and competes with the decays coming from $\Gamma_\phi\rho_\phi$. For higher values of $\kappa$, the contribution of the viscosity is almost neglected concerning the decay term and, therefore, the case with and without bulk viscosity are similar. In the lower values of $\kappa$, the effects of viscosity are dominant over $\Gamma_\phi\rho_\phi$ and show a different behavior from the NSC, explaining why for a certain value of $T_\text{end}$ the value of $\kappa$ is not relevant. This fact can be appreciated when the blue line crosses the red area in which $\rho_\phi<\rho_\gamma$ in the NSC, meaning that the entropy injection from the $\phi$ decays is neglected. Nevertheless, the viscosity included in the new state makes significant imprints in the entropy injection for radiation. The latter can be understood considering that the right hand side of Eq. \eqref{eqphibulk} can be rearranged as
\begin{equation}
    \rho_\phi\left(-\Gamma_\phi+\frac{3\hat{\xi_0}H^2\rho_\phi^{l-1}}{M_p^{4m-3}}\right)=\rho_\phi\left(-\Gamma_\phi+\nu_\phi\right), \label{arrengement}
\end{equation}
where we defined $\nu_\phi$ as the viscous term on the left hand side of Eq. (\ref{arrengement}). This helps us to visualize the dominance between the decay and the viscosity. The Fig. \ref{figgammanu} depicts the behavior of these two quantities for $\omega=0$ and $\hat{\xi_0}=10^{-2}$. The figure shows a benchmark for three points of the form $\left(T_\text{end},\,\kappa\right)$: RI for $(10^{-2}\,\text{GeV},\,10^{-3})$ (dashed-dotted line), RII for $(0.1\,\text{GeV},\,0.1)$ (dashed line), and RIII for $(2\,\text{GeV},\,100)$ (solid line). The blue color palette represent the evolution of the $\nu_\phi$ term and the horizontal lines (red color palette) represents the $\Gamma_\phi$ term. The vertical lines (purple color palette) are the values of $T_\text{end}$ for the three points mentioned above. This illustrates why the parameter space is similar for higher values of $\kappa$, because for a wide range of temperatures, the term $\Gamma_\phi$ dominates over $\nu_\phi$. Meanwhile, the value of $\kappa$ and $T_\text{end}$ are diminishing, there is a significant parameter space in which $\nu_\phi$ dominates over $\Gamma_\phi$. Moreover, to even lower values of $\kappa$ for $T_\text{end}=3\times 10^{-3}$ GeV, the solutions are very similar to the point evaluated in RI, explaining the large values of $\kappa$ that reproduce the DM relic density for almost the same values of $T_\text{end}$. Note, from Figs. \ref{figkappaTendw0}, \ref{figkappaTendw-25}, and \ref{figkappaTendw25}, that this feature described applied merely when the RI exist ($\omega\leq1/3$).

As it was shown, the inclusion of viscosity changes the evolution of the energy density. This implies an increment in the entropy injection to radiation leading into lower values of the parameter space to search the DM relic density. This behavior can also be seen for different values of the barotropic index $\omega$. For example, in Figs. \ref{figkappaTendw-25} and \ref{figkappaTendw25}, we shown the parameter space of the model that reproduces the current DM relic density for the NSC with (blue line) and without (red line) bulk viscosity for $m_\chi=100$ GeV and $\langle\sigma v\rangle=10^{-11}$ GeV$^{-2}$. In the case of $\omega=-2/5$ (Fig. \ref{figkappaTendw-25}), the parameter space is highly different between both scenarios because the viscous term dominate the behaviour of the NSC earlier, observing that for the certain value $T_\text{end}\sim8\times 10^{-1}$ GeV there is a vast space of $\kappa$-values that reproduce the current DM relic density. On the other hand, for $\omega=2/5$ (Fig. \ref{figkappaTendw25}), we do not see the aforementioned behaviour since the RI does not exists. However, there is almost one order of magnitude of difference between both scenarios, allowing lower values of $\kappa$ that can reproduce the current DM relic density.

\begin{figure}
\centering
\includegraphics[width=0.4\textwidth]{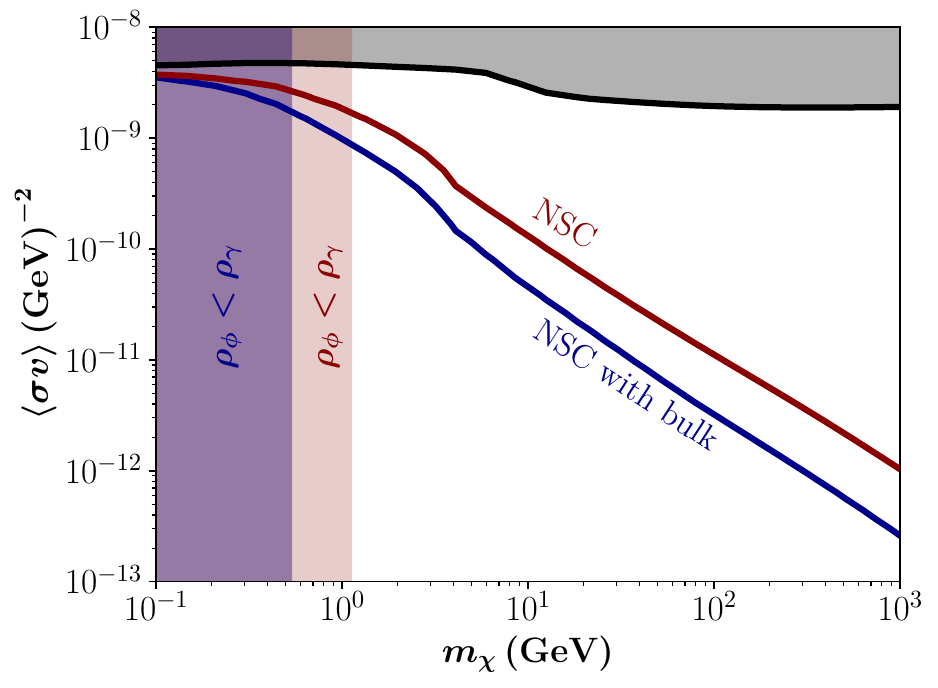}
\caption{Parameter space $\left(m_\chi,\langle\sigma v\rangle\right)$ for the DM with $\kappa=10^{-2}$, $T_{end}=7\times 10^{-3}$ GeV, $\omega=0$, and $\hat{\xi_0}=10^{-2}$. The black, red, and blue lines correspond to the parameters that reproduce the observed DM relic density in the $\Lambda$CDM model, the classical NSC, and the NSC with bulk viscosity, respectively. The $\Lambda$CDM model has a total thermal averaged annihilation cross-section $\langle\sigma v\rangle_0 = \text{few}\times 10^{-9}$ GeV$^{-2}$, for WIMPs candidates, with the grey zone corresponding to parameters that are excluded even for the NSC scenarios. The red and blue zones (NSC and NSC with bulk viscosity, respectively) correspond to the parameters space in which the energy density of the new field $(\phi)$ is lower than radiation and, therefore, are close to the $\Lambda$CDM case.}
\label{figmsigma}
\end{figure}

Finally, in Fig. \ref{figmsigma}, we present the parameter space for the WIMP DM candidate, namely, its mass $(m_\chi)$ and total thermal averaged annihilation cross-section $(\langle\sigma v\rangle)$, for the $\Lambda$CDM model (black line), the classical NSC (red line), and the NSC with bulk viscosity (blue line). We consider the particular case where $\omega=0$, $\kappa=10^{-2}$, and $T_\text{end}=7\times 10^{-3}$ GeV. Note that, if the DM is established in the $\Lambda$CDM model, then the total thermal averaged annihilation cross-section for the candidate must be $\langle\sigma v\rangle_0=\text{few}\, \times 10^{-9}$ GeV$^{-2}$ in the range of mass considered. However, both NSC scenarios reach this limit only when the DM mass is decreasing. Therefore, values of the total thermal averaged annihilation cross-section higher than $\langle\sigma v\rangle_0$ are not allowed and are represented in the gray zone. The blue and red zones represent the condition $\rho_\phi<\rho_\gamma$ for which the parameters that reproduce the DM relic density go closer to the $\Lambda$CDM case for the NSC with and without bulk viscosity, respectively. This behaviour is due to the low quantity of entropy injected to radiation. It is important to highlight that the inclusion of the bulk viscosity, as it was shown before, leads to a down displacement in the values for the DM parameters such as  $(\kappa,\,T_\text{end})$ (see Figs. \ref{figkappaTendw0} - \ref{figkappaTendw25}).

\subsection{\label{sec:Parameters} Parameter space for dark matter}
We have already presented the differences between the NSC with and without bulk viscosity, studying the entropy injection, the DM production, and the imprints in the parameters space to obtain the current DM relic density. Now, we are interested in the study of the NSC with bulk viscosity in two perspectives: (i) if we detect a DM signal with specific parameters $(m_\chi,\,\langle\sigma v\rangle)$, which cosmological model could adjust those parameters? and (ii) for a specific model benchmark, which are the DM parameters that could reproduce the current observable relic density? 

\begin{figure}
\centering
\includegraphics[width=0.4\textwidth]{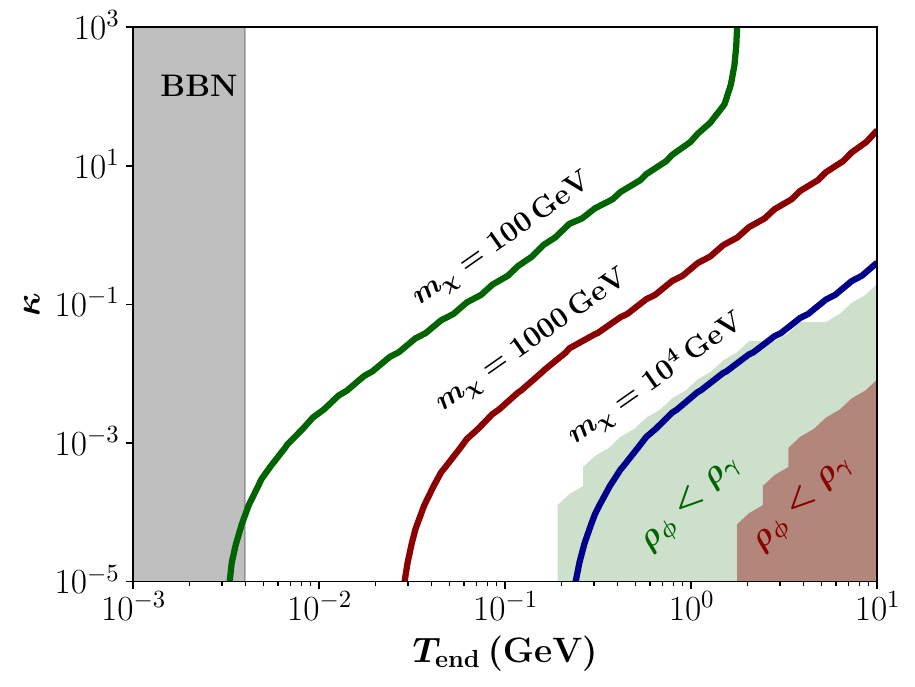}
\caption{Parameter space $(T_\text{end},\kappa)$ for the model with $\omega=0$, $\hat{\xi_0}=10^{-2}$, and $\langle\sigma v\rangle=10^{-11}$ GeV$^{-2}$. The green, red, and blue lines correspond to the parameter space that reproduces the DM relic density for $m_\chi=100 ,\, 1000,\,\, \text{and}\,\, 10^4$ GeV, respectively. The green and red areas correspond to the parameters space where $\rho_\phi<\rho_\gamma$, for all time, for $m_\chi=100, \,\, \text{and}\,\, 1000$ GeV, respectively. The area for $m_\chi=10^4$ GeV is shifted to lower values of $\kappa$ and higher values of $T_\text{end}$. The variation of DM mass shifts the curves to the right when the value of $m_\chi$ increases (and vice-versa). The grey zone on the left corresponds to the BBN epoch that starts at $T_\text{BBN}\sim 4\times 10^{-3}$ GeV.}
\label{figvarmdm}
\end{figure}

\begin{figure}
\centering
\includegraphics[width=0.4\textwidth]{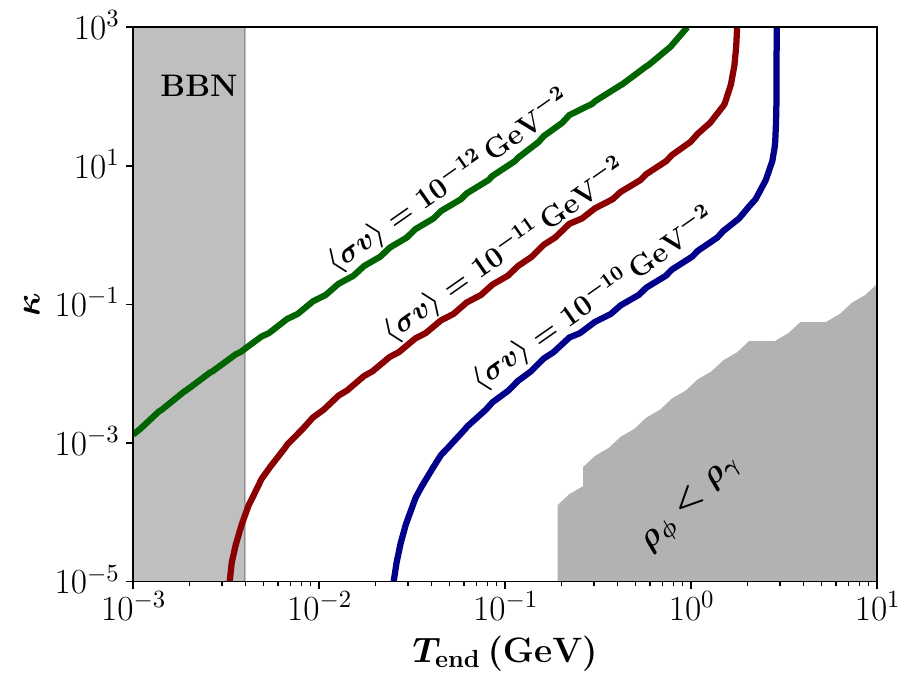}
\caption{Parameter space $(T_\text{end},\kappa)$ for the model with $\omega=0$, $\hat{\xi_0}=10^{-2}$, and $m_\chi=100$ GeV. The blue, red, and green lines correspond to the parameter space that reproduces the DM relic density for $\langle\sigma v\rangle = 10^{-10},\, 10^{-11},\,\, \text{and}\,\, 10^{-12}$ GeV$^{-2}$, respectively. The gray area correspond to the parameters space where $\rho_\phi<\rho_\gamma$, for all time, which is the same in the three cases. The variation of the total thermal averaged annihilation cross-section shifts the curves to downward when the value of $\langle\sigma v\rangle$ increases (and vice-versa). The grey zone on the left corresponds to the BBN epoch that starts at $T_\text{BBN}\sim 4\times 10^{-3}$ GeV.}
\label{figvarsv}
\end{figure}

\begin{figure}
\centering
\includegraphics[width=0.4\textwidth]{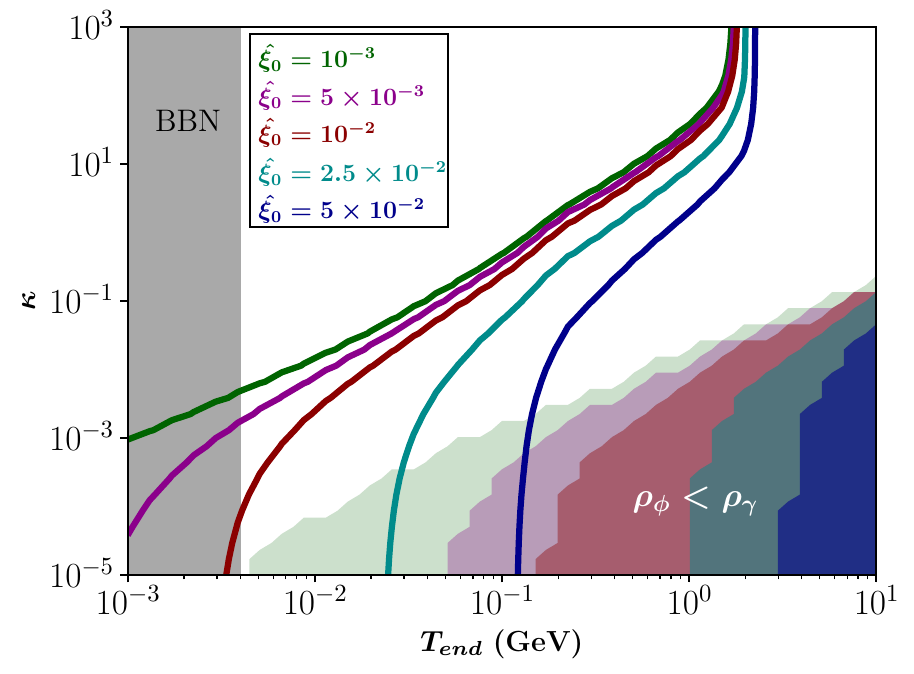}
\caption{Parameter space $(T_\text{end},\kappa)$ for the model with $\omega=0$, $\langle\sigma v\rangle=10^{-11}$ GeV$^{-2}$, and $m_\chi=100$ GeV. The green, purple, red, cyan, and blue lines correspond to the parameter space that reproduces the DM relic density for $\hat{\xi}_0=10^{-3}$, $5\times10^{-3}$, $10^{-2}$, $2.5\times10^{-2}$, and $5\times10^{-2}$, respectively. The green, purple, red, cyan, and blue areas correspond to the parameters space where $\rho_\phi<\rho_\gamma$, for all time, for the same values of $\hat{\xi}_0$, respectively. Note that higher values of $\hat{\xi}_0$ lean the curves to a region in which the model is independent of the value for $\kappa$, meanwhile lower values tends to the classical NSC. The grey zone on the left corresponds to the BBN epoch that starts at $T_\text{BBN}\sim 4\times 10^{-3}$ GeV.}
\label{figvarxi0}
\end{figure}

The first perspective is illustrated in Figs. \ref{figvarmdm}, \ref{figvarsv}, and \ref{figvarxi0}. In particular, in Fig. \ref{figvarmdm}, we depict the parameter space $(T_\text{end},\kappa)$ that reproduces the current DM for $\omega=0$ and $\langle\sigma v\rangle=10^{-11}$ GeV$^{-2}$, considering three particular cases, namely, $m_\chi=100 ,\, 1000\,\, \text{and}\,\, 10^4$ GeV. The most important conclusion is that the curves of parameters space allowed in ($T_\text{end}$, $\kappa$) are shifted to the right (and slightly down) when the DM mass is higher (and vice-versa), which is also applied to the restricted areas $\rho_\phi<\rho_\gamma$. On the other hand, in Fig. \ref{figvarsv}, we depict the same parameter space $(T_\text{end},\kappa)$ but for a fixed DM mass given by $m_\chi=100$ GeV, considering three particular cases, namely, $\langle\sigma v\rangle=10^{-10}$, $10^{-11}$, and $10^{-12}$ GeV$^{-2}$. Again, we can see a shift of the curves when is varied the values of the total thermal averaged annihilation cross-section $\langle\sigma v\rangle$ that generate the observed DM abundance. In particular, the curves are displaced downward (and slightly to the right) if the value of $\langle\sigma v\rangle$ increases (and vice-versa). Note that, in this case, the restricted zone $\rho_\phi<\rho_\gamma$ is not affected, taking the same values for all the variations of $\langle\sigma v\rangle$. It is important to note that the region in which the model becomes independent of the values for $\kappa$ (in the particular case when $\omega<1/3$) can be displaced to avoid the BBN epoch, as it is possible to see in Fig. \ref{figkappaTendw0}. In particular, the latter can be done for a fixed $\langle\sigma v\rangle=10^{-11}$ GeV$^{-2}$ value and a DM mass in the range $m_\chi>100$ GeV; or for a fixed DM mass $m_\chi=100$ GeV value and a total thermal averaged annihilation cross-section in the range $\langle\sigma v\rangle_0>\langle\sigma v\rangle>10^{-11}$. In general, higher values of $m_\chi$ combined with lower values of $\langle\sigma v\rangle$ would open this window to explore. Finally, in Fig. \ref{figvarxi0}, we depict the parameter space $\left(T_\text{end},\kappa\right)$ for the model that reproduces the current DM relic density for $\omega=0$, $m_\chi=100$ GeV, and $\langle\sigma v\rangle=10^{-11}$ GeV$^{-2}$, considering five cases, namely, $\hat{\xi}_0=10^{-3}$, $5\times10^{-3}$, $10^{-2}$, $2.5\times10^{-2}$, and $5\times10^{-2}$. For higher values of $\kappa$ (RIII), there are no significant differences among the curves, meanwhile, for lower values of $\kappa$ (RII and I), the differences are significant when the value of $\hat{\xi_0}$ increases. Also, when $\hat{\xi_0}\to 0$, the curves tend to the classical NSC scenario because the dissipation is negligible and Eq. \eqref{eqphibulk} reduces to Eq. \eqref{phi}. On the other hand, higher values of viscosity make more prominent the below curvature, generating a shorter range of $T_\text{end}$ for a large range of $\kappa$.

\begin{figure}
\centering
\includegraphics[width=0.4\textwidth]{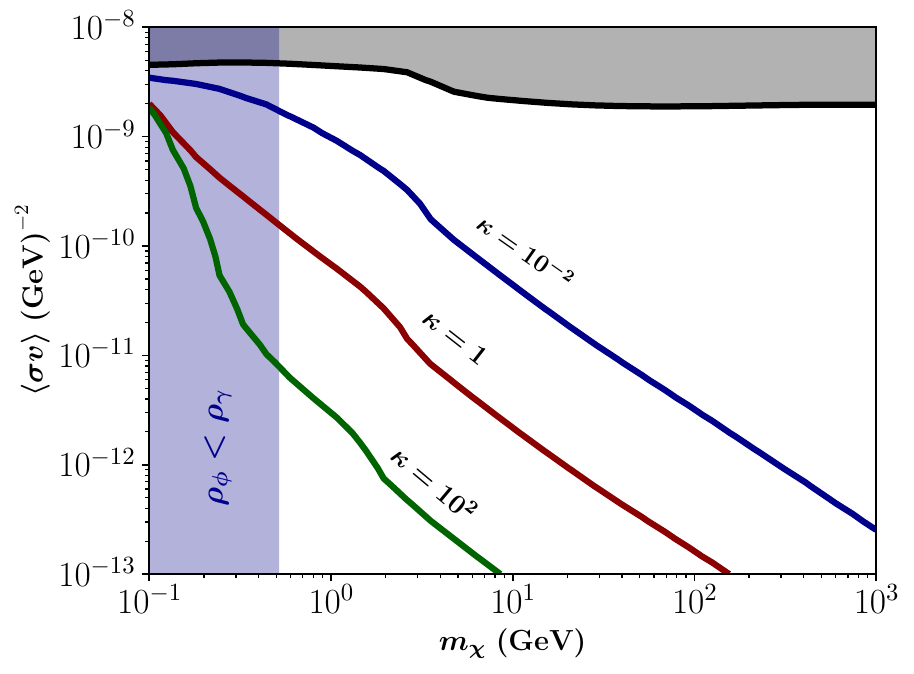}
\caption{Parameter space $(m_\chi,\langle\sigma v\rangle)$ for the DM candidate with $\omega=0$, $\hat{\xi_0}=10^{-2}$, and $T_\text{end}=7\times 10^{-3}$ GeV. The green, red, and blue lines correspond to the parameter space that reproduce the DM relic density for $\kappa = 10^2,\, 1,\, \text{and}\, 10^{-2}$, respectively. The grey area represents the parameters not allowed in the $\Lambda$CDM model and the black line corresponds to the DM parameters that reproduces its current density in the same model at $\langle\sigma v\rangle_0=\text{few}\times 10^{-9}$ GeV$^{-2}$. The blue zone are the parameters space where $\rho_\phi<\rho_\gamma$, for all time, with $\kappa=10^{-2}$. Note that higher values of $\kappa$ shift this area to the left until $\kappa=1$, while for $\kappa>1$ this area does not exist. Also, an increment in the $\kappa$-values allows to achieve lower values of $\langle\sigma v\rangle$.}
\label{figvarkappa}
\end{figure}

\begin{figure}
\centering
\includegraphics[width=0.4\textwidth]{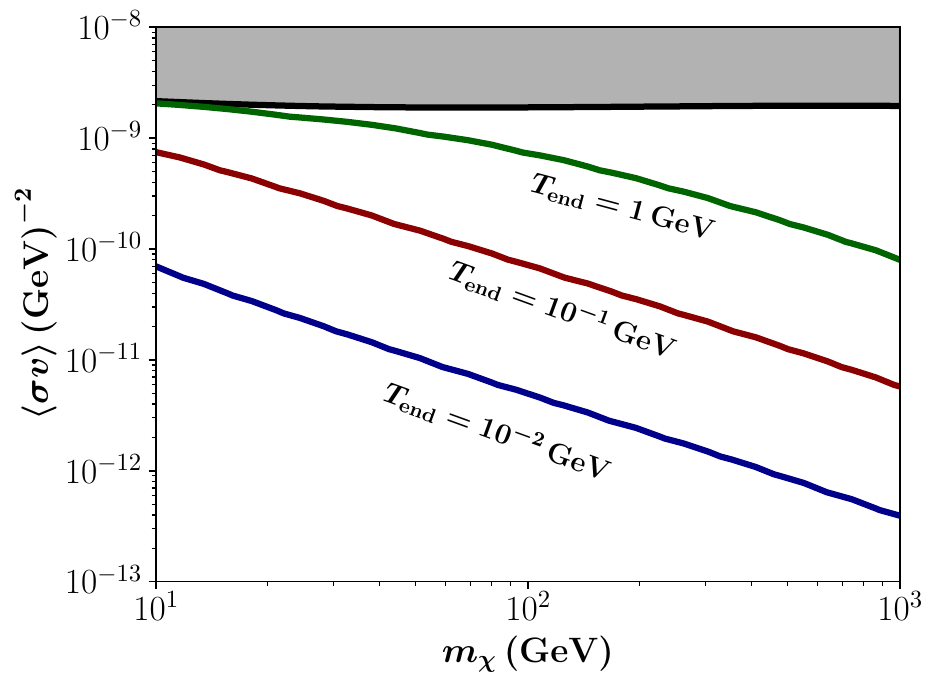}
\caption{Parameter space $(m_\chi,\langle\sigma v\rangle)$ for the DM candidate with $\omega=0$, $\hat{\xi_0}=10^{-2}$, and $\kappa=10^{-2}$. The green, red, and blue lines correspond to the parameter space that reproduce the DM relic density for $T_\text{end}=1$, $10^{-1}$, and $10^{-2}$ GeV, respectively. The grey area represents the parameters not allowed in the $\Lambda$CDM model and the black line corresponds to the DM parameters that reproduces its current density in the same model at $\langle\sigma v\rangle_0=\text{few}\times 10^{-9}$ GeV$^{-2}$. Note that an increment in $T_\text{end}$-values tends the NSC with bulk viscosity to the $\Lambda$CDM scenario due to the rapidly decays of $\phi$.}
\label{figvartend}
\end{figure}

\begin{figure}
\centering
\includegraphics[width=0.4\textwidth]{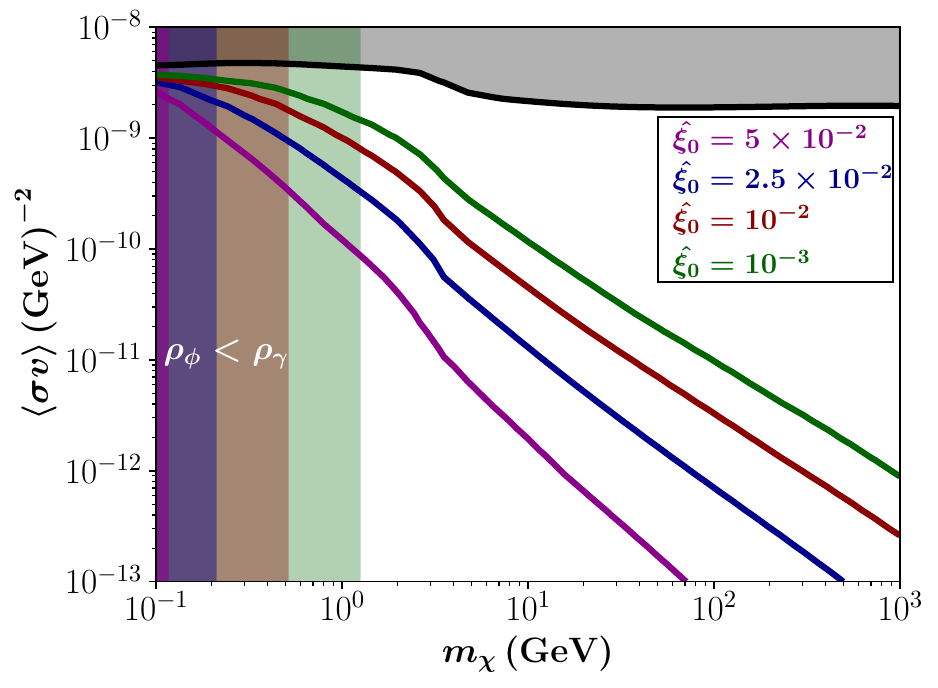}
\caption{Parameter space $(m_\chi,\langle\sigma v\rangle)$ for the DM candidate with $\omega=0$, $T_\text{end}=7\times10^{-3}$ GeV, and $\kappa=10^{-2}$. The green, red, blue, and purple lines correspond to the parameter space that reproduces the DM relic density for $\hat{\xi}_0=5\times10^{-2}$, $2.5\times10^{-2}$, $10^{-2}$, and $10^{-3}$, respectively. The grey area represents the parameters not allowed in the $\Lambda$CDM model and the black line corresponds to the DM parameters that reproduce its current density in the same model at $\langle\sigma v\rangle_0=\text{few}\times 10^{-9}$ GeV$^{-2}$. The colored areas correspond to $\rho_\phi<\rho_\gamma$ for the same values of $\hat{\xi}_{0}$, respectively. Note that for lower values of $\hat{\xi}_0$ the model tends to the NSC scenario without bulk viscosity, meanwhile, higher values of $\hat{\xi}_0$ shift the DM parameters that can reproduce the DM relic density to the left.}
\label{figvarxi0mdm}
\end{figure}

The second perspective is illustrated in Figs. \ref{figvarkappa}, \ref{figvartend}, and \ref{figvarxi0mdm}, where we study the free parameters of the DM for different elections of $\kappa$, $T_\text{end}$, and $\hat{\xi}_0$, respectively. In the figures, the grey zone corresponds to the DM parameter space not allowed in the $\Lambda$CDM model and the black line to the parameter space that reproduces the current DM relic density in the same model. In particular, in Fig. \ref{figvarkappa}, we depict the parameter space ($m_\chi$, $\langle\sigma v\rangle$) that reproduce the current DM relic density for $\omega=0$, $\hat{\xi_0}=10^{-2}$, and $T_\text{end}=7\times 10^{-3}$ GeV, considering three particular cases, namely, $\kappa=10^2$, $1$, and $10^{-2}$. From the figure, we can see that only for $\kappa<1$ exists the region in which $\rho_\phi<\rho_\gamma$, for all time. If the values of $\kappa$ increase, then the parameter space allowed to shift the total thermal averaged annihilation cross-sections to lower values. Also, when $\kappa\ll1$, we recover the $\Lambda$CDM scenario. On the other hand, in Fig. \ref{figvartend}, we depict the parameter space ($m_\chi$, $\langle\sigma v\rangle$) that reproduce the current DM relic density for $\omega=0$, $\hat{\xi_0}=10^{-2}$, and $\kappa=10^{-2}$, considering three particular cases, namely, $T_\text{end}=10^{-2}$, $10^{-1}$, and $1$ GeV. From the figure, we can see that higher values of $T_\text{end}$ tend the NSC with bulk viscosity to the $\Lambda$CDM scenario, since the $\phi$ state decays rapidly and there is no significant entropy injection, meanwhile, lower values of $T_\text{end}$ allow a large range of $\langle\sigma v\rangle$. Finally, in Fig. \ref{figvarxi0mdm}, we depict the parameter space ($m_\chi$, $\langle\sigma v\rangle$) that reproduces the current DM relic density for $\omega=0$, $\kappa=10^{-2}$, and $T_\text{end}=7\times10^{-3}$ GeV, considering four particular cases, namely, $\hat{\xi}_0=5\times10^{-2}$, $2.5\times10^{-2}$, $10^{-2}$, and $10^{-3}$. From this figure we can see that for lower values of $\hat{\xi}_0$ the model tends to the NSC scenario without bulk viscosity (see also Fig. \ref{figvarxi0}), meanwhile, higher values of $\hat{\xi_0}$ shift the DM parameters that can reproduce the relic density to the left, splitting the NSC with bulk viscosity from the classical NSC case and allowing lower values in $\langle\sigma v\rangle$. Again, the colored zones represent the cases when $\rho_\phi<\rho_\gamma$ at any time. Therefore, the dissipation of the $\phi$ field gives us new zones to search for the WIMPs candidates. An important result can be noticed for different values of $\hat{\xi}_0$. These values generate a displacement in the parameter space which translates into that different values of $\hat{\xi}_{0}$ could reproduce different NSCs scenarios without bulk viscosity, i.e, a specific value of $\hat{\xi}_{0}$ and $\omega$ can match the same parameter space in a classical NSC with different $\omega$. From this analysis we can see that in the bulk viscous non-standard cosmologies the cases where $\omega>-1$ but close to $-1$ must be treated carefully. This is because we can obtain a fluid that never decay, due to the effects of the viscosity, obtaining a behaviour similar to the standard NSC for a field with $\omega\leq -1$.

\begin{figure}
\centering
\includegraphics[width=0.4\textwidth]{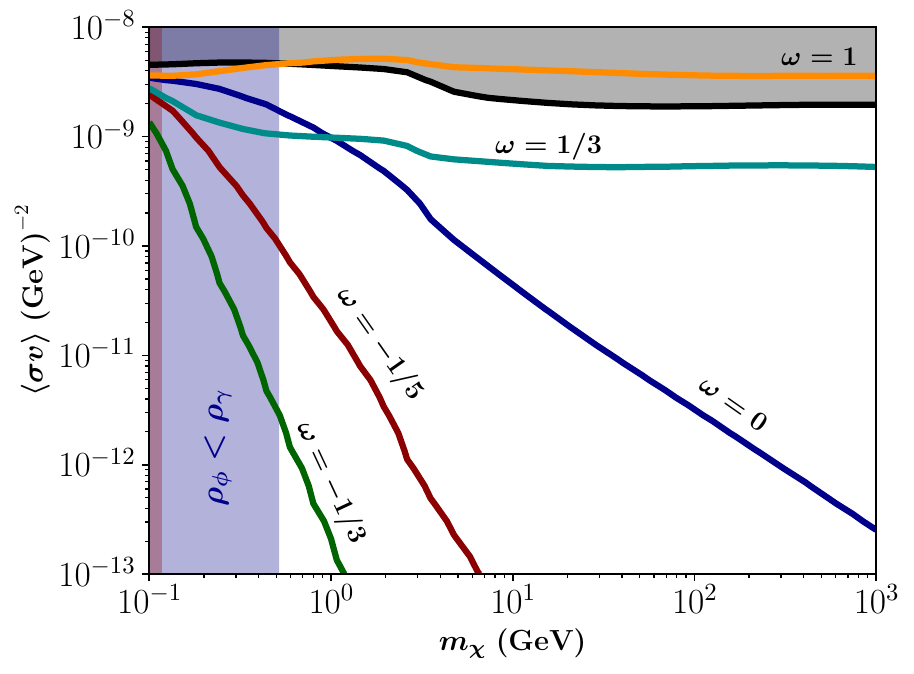}
\caption{Parameter space $(m_\chi,\langle\sigma v\rangle)$ for the DM candidate with $T_\text{end}=7\times10^{-3}$ GeV and $\hat{\xi}_0=10^{-2}$. The green, red, blue, cyan, and orange lines correspond to the parameter space that reproduces the DM relic density considering $\kappa=10^{-2}$ for the barotropic index $\omega=-1/3$, $-1/5$, and $0$; $\kappa=10^2$ for $\omega=1/3$; and $\kappa=10^4$ for $\omega=1$, respectively. The grey area represents the parameters not allowed in the $\Lambda$CDM model and the black line corresponds to the DM parameters that reproduce its current density in the same model at $\langle\sigma v\rangle_0=\text{few}\times 10^{-9}$ GeV$^{-2}$. The blue and red zones are the parameters space where $\rho_\phi<\rho_\gamma$, for all the time, when $\omega=0$ and $-1/5$, respectively. Note that the Kination case ($\omega=1$) allows a total thermal averaged annihilation cross-sections higher than  $\langle\sigma v\rangle_0$ for almost all the range of DM mass shown.}
\label{figvaromega}
\end{figure}

For a further comparison, in Fig. \ref{figvaromega}, we consider different kinds of fluids for $T_\text{end}=7\times 10^{-3}$ GeV and $\hat{\xi}_0=10^{-2}$. To solve the differential equations, we consider the value $\kappa=10^{-2}$ for the barotropic index $\omega=-1/3$, $-1/5$, and $0$; $\kappa=10^2$ for $\omega=1/3$; and $\kappa=10^4$ for $\omega=1$. The consideration of different values in $\kappa$ is related to the evolution of the fluid itself, i.e., fluids with $\omega>1/3$ diluted rapidly compared to radiation and, therefore, need higher initial energy to generate an effective entropy injection to radiation before the decay of $\phi$. From the figure we can see that, for the curves with $\omega<0$, their slopes tend to lean downward. On the other hand, the curves with $\omega>0$ tend to lean their slopes upward. The case with $\omega=1$ must be analyzed carefully because it enters into the forbidden $\Lambda$CDM zone, which translates into values of $\langle\sigma v\rangle$ slightly higher than $\langle\sigma v\rangle_0$ for a DM range mass of $10^{-1}$ GeV $\leq m_\chi\leq 10^3$ GeV.

\section{\label{sec:Conclusions}Conclusions}
In this paper, we explored an extension of the classical NSC scenario in which the new field $\phi$, which interacts with the radiation component in the early universe, experiences dissipative processes in the form of a bulk viscosity. Working in the framework of Eckart's theory, we studied the difference between both scenarios, considering a bulk viscosity proportional to the energy density of the field according to the expression $\xi=\xi\rho_{\phi}^{1/2}$. In addition to being one of the most studied, this parameterization has the characteristic that, when the field $\phi$ fully decays in SM plasma, the dissipation becomes negligible and we recover the $\Lambda$CDM model without viscosity. Following this line, in the case that DM is discovered with its physics parameters ($m_\chi$,$\langle\sigma v\rangle$) reconstructed, it is imperative to determine if those parameters are in agreement with the $\Lambda$CDM model or not. Hence, the inclusion of this novel NSC scenario brings to life parameters for WIMPs DM candidates that were discarded in the $\Lambda$CDM model and in the classical NSC scenario, obtaining new regions or re-open windows to search them. We study this new NSC assuming the most studied interacting term of the form $\Gamma_{\phi}\rho_{\phi}$, searching for the parameter space for DM production that leads the current observable relic density.

As it was shown in Figs. \ref{figkappaTendw0}, \ref{figkappaTendw-25}, and \ref{figkappaTendw25}, the model parameters that reproduce the right abundance of DM relic are very close to higher values of $\kappa$ (RIII). On the contrary, when $\kappa$ decreases, the case with viscosity shows clear differences, as the independence in $\kappa$-values of the model in RI, similarly as in RIII. This behavior is merely by the inclusion of the viscosity as it was shown in Fig. \ref{figvarxi0}, in which lower values in $\hat{\xi}_0$ tend to reproduce the NSC scenario without bulk viscosity and higher values of $\hat{\xi}_0$ generate the independent zone in $\kappa$ (for RI) sooner. The variation of DM mass or $\langle\sigma v\rangle$ shifts the parameters to the left/right or up/down when the parameters mentioned are lower/higher, respectively. On the other hand, when the DM parameters are explored for specific benchmarks of the model, it can be seen from Fig. \ref{figmsigma} that present new zones to obtain the current relic density, giving the possibility to reach lower values of $\langle\sigma v\rangle$ for the range of mass studied. Nevertheless, for lower values of DM mass, the NSC with and without bulk viscosity are similar. Fig. \ref{figvarxi0mdm} also shows that for lower values of $\hat{\xi}_0$, the parameters are similar to the classical NSC case, meanwhile $\hat{\xi}_0$ is higher, the slope in the curves of parameters is more pronounce to lower values of $\langle\sigma v\rangle$, i.e, the inclusion of higher values of $\hat{\xi}_0$ provides lowers values for the total thermal averaged annihilation cross-section. Finally the variation of the model parameters $\kappa$/$T_\text{end}$ shifts the current DM relic density to left/right or up/down when the parameters mentioned are higher/lower, respectively.

Therefore, this paper is a further step in the study of WIMPs as DM candidates and a first approximation to highlighting the imprints that the bulk viscosity can leave in these particles and their relic density in the early universe through a NSC scenario.

\section*{\label{sec:Acknowledgments}Acknowledgments}
E.G. was funded by Vicerrectoría de Investigación y Desarrollo Tecnológico (VRIDT) at Universidad Católica del Norte (UCN) through Proyecto de Investigación Pro Fondecyt 2023, Resolución VRIDT N°076/2023. He also acknowledges the scientific support of Núcleo de Investigación No. 7 UCN-VRIDT 076/2020, Núcleo de Modelación y Simulación Científica (NMSC).

\newpage
\bibliographystyle{apsrev4-2}
\bibliography{bibliography}
\end{document}